\documentclass[ usenatbib]{emulateapj}

\usepackage{graphicx}
\usepackage{amsmath}
\usepackage{natbib}
\usepackage{paralist}
\usepackage{nicefrac}
\usepackage{ctable}
\usepackage[normalem]{ulem}

\begin{document}

\title{Mass Measurements of Black Holes in X-Ray Transients: Is There A Mass Gap?}
\subjectheadings{black hole physics --- X-rays: binaries}

\author{Laura Kreidberg$^{1,2}$, Charles D. Bailyn$^{1}$, Will M. Farr$^{3,4}$, Vicky Kalogera$^{3}$}
\affiliation{$^{1}${Department of Astronomy, Yale University, New Haven, CT 06520 USA}, \\
$^{2}${Department of Astronomy \& Astrophysics, University of Chicago, Chicago, IL 60637, USA}, \\
$^{3}${Department of Physics and Astronomy \& Center for Interdisciplinary Exploration and Research in Astrophysics (CIERA),}\\
{ Northwestern University, Evanston, IL 60208, USA},\\
$^{4}${CIERA Postdoctoral Fellow}}

\begin{abstract}
We explore possible systematic errors in the mass measurements of stellar mass black holes.  We find that significant errors can arise from the assumption of zero or constant emission from the accretion flow, which is commonly used when determining orbital inclination by modelling ellipsoidal variations.  For A0620-00, the system with the best available data, we show that typical data sets and analysis procedures can lead to systematic underestimates of the inclination by ten degrees or more.  A careful examination of the available data for the 15 other X-ray transients with low-mass donors suggests that this effect may significantly reduce the black hole mass estimates in several other cases, most notably that of GRO J0422+32.  With these revisions, our analysis of the black hole mass distribution 
in soft X-ray transients does not suggest any ``mass gap" between the low end of the distribution and the maximum theoretical neutron star mass, as has been identified in previous studies.
Nevertheless, we find that the mass distribution retains other previously identified characteristics, namely a peak around $8M_{\odot }$, a paucity of sources with masses below $5M_{\odot }$, and a sharp drop-off above $10M_{\odot }$.
\end{abstract}

\section{Introduction}
Soft X-ray transients (SXTs) provide some of the strongest evidence for the existence of stellar mass black holes.  In these systems, Eddington limited X-ray outbursts occur over timescales of weeks to months, followed by years to decades of X-ray quiescence \citep{chen97}.  During quiescence, the optical light from the system is dominated by the companion star.  Classic binary star analysis techniques can be used to determine the orbit of the companion star in quiescence, and thus the orbital parameters of the entire system, including precise determinations of the mass of the compact object \citep{orosz03}. Many of these mass determinations are greater than the theoretical upper bound of neutron stars ($\approx 3 M_\odot$), and thus the compact objects are understood to be black holes.  
These black hole SXTs (BHSXTs) comprise most of the dynamically confirmed stellar mass black holes.

Mass measurements of these systems have been used to explore the distribution of black hole masses in X-ray binaries \citep{bailyn98,ozel10,farr10}.  In these studies, published values for the orbital parameters have been used as inputs for a Bayesian analysis of the mass distribution of the compact objects.  In all of these analyses, a significant mass gap between the maximum neutron star mass ($3M_{\odot }$) and the low end of the black hole mass distribution ($\lesssim 5M_{\sun }$) has been identified.  This is a curious result, since one would ordinarily expect that the mass distribution of black holes would be weighted toward the low end, as is the mass distribution of pre-supernova massive stars. Several theories have been proposed
to explain the evolution of massive stars in binary systems and how the resultant supernova explosions might result in such a mass gap \citep{brown01, belczynski11}. 

Recently, however, the accuracy of the mass measurements for individual objects has implicitly
been questioned.  Cantrell et al. (2010, hereafter C10) have reanalyzed all extant data for the prototypical BHSXT A0620-00.  Prior to C10,
published values for the orbital inclination ranged from $i=37^\circ$, implying
a black hole mass of $16 M_{\odot }$ \citep{shahbaz94} to $i=74^\circ$, implying a black hole mass of $4 M_{\odot }$ \citep{froning01} with several intermediate
values \citep{haswell93,gelino01}. C10 find that this wide range of incompatible results can be reconciled by more careful modelling of the ellipsoidal light curves of
the companion star.  In particular, C10 cull the data to include only data in the ``passive'' state \citep{cantrell08}, in which the lightcurves do not exhibit short-term non-ellipsoidal variability.  They fit
the lightcurves with a model that includes variable disk light and a hotspot that is allowed to vary in position and brightness.  
In this way C10 find a consistent value of the orbital inclination, $i = 51.0 \pm 0.9^\circ$, which implies $m_\mathrm{BH} = 6.6 \pm 0.25 M_\odot$.

It is becoming increasingly clear that other BHSXTs have
active/passive state changes similar to A0620-00, and may have variable hotspots \citep[e.g.,][]{macdonald11}.  However, the orbital parameter estimates in the literature generally do not take these effects into account. Thus it is possible that, like A0620-00, many of the BHSXTs may have mass estimates that are inaccurate by considerable amounts.  In this paper
we examine the systematic errors introduced by the presence of nonstellar flux
in the ellipsoidal lightcurves of BHSXTs.  We find that there is likely a bias toward mass estimates that are higher than the true mass of the compact object.  
We find that the ``mass gap'' identified in previous work \citep{bailyn98, ozel10,farr10} on the mass distribution of BHSXTs can be accounted for by this systematic effect.

The outline of the paper is as follows.
In $\S 2$ we summarize the primary sources of systematic uncertainty in black hole mass measurements.  We quantify these uncertainties in $\S 3$ for A0620-00.
In $\S 4$ we generalize the error estimates obtained for A0620-00 to other systems.  $\S 5$ contains a re-evaluation of the mass estimates for 16 BHSXTs.  We use these
revised mass estimates to analyze the mass distribution in $\S 6$, and in $\S 7$ we conclude.

\section{Sources of Systematic Error in Mass Determination}
The mass $m_\mathrm{BH}$ of the black hole in a BHSXT is determined by three parameters:  the mass ratio of the secondary star
to the black hole, $q \equiv m_*/m_\mathrm{BH}$, the mass function $f$, and the orbital inclination $i$.  
Written in terms of these parameters, the black hole mass is
\begin{equation}
m_\mathrm{BH} = \dfrac{f(1 + q)^2}{\sin^3 i}.
\end{equation} 
The error on the mass measurement thus depends on the error in $f$, $q$, and $i$.

The mass function $f$ generally contributes little to the systematic error on black hole mass.
It is a function of the orbital period $P$ and semi-amplitude of the radial velocity curve $K$, $f = PK^3/(2\pi G)$.
For most cases, $P$ is measured to sub-percent precision using photometric observations that span many periods.
The radial velocity curve amplitude is typically known to within $10\%$, with precision mainly limited by the resolution of the spectroscopy.
Neither measurement is strongly affected by systematic bias.
A narrow Gaussian is thus a good approximation of the error on $f$, which 
introduces a small random error on $m_\mathrm{BH}$.

The mass ratio $q$ is inferred from the rotational broadening of spectral lines.
By inspection of equation (1), we see that $m_\mathrm{BH}$ is relatively insensitive to the mass ratio for $q \ll 1$.
In 11 of the 16 systems in our sample, $q < 0.15$ (see Table 2).  For three systems with larger $q$,
the mass ratio is well-constrained.  The remaining two systems have only an upper limit, $q < 0.5$.
Thus in all but these two systems the uncertainty in the measured mass ratio has a small impact on $m_\mathrm{BH}$.

By far the largest source of systematic error is the orbital inclination.  
Inclination is typically measured by analyzing ellipsoidal variability in the observed photometric lightcurve.  The origin of this variability is
gravitational distortion of the companion star.  The star fills its Roche lobe, so as it orbits the black hole, the projected 
surface area and average temperature along the observer's line of sight is not constant.  This gives rise to characteristic double peaks in the lightcurve, known as ellipsoidal variations, 
whose amplitude depends on the inclination.  The largest amplitude occurs for systems edge-on to our line of sight ($i = 90^\circ$), 
because that geometry maximizes the changes in projected surface area of the star with orbital phase.  By contrast, no ellipsoidal variations can be detected for face-on systems ($i = 0^\circ$). 
One can therefore determine $i$ by modeling ellipsoidal variability in the observed photometric lightcurve.

The simplest model of ellipsoidal variability is just a simulated Roche-lobe filling star.  This model, which we will call the `star-only' model,
is commonly employed in the literature under the assumption that nonstellar sources of light are negligible \citep[e.g.][]{martin95,greene01, gelino03}.  
It is not unusual, however, for nonstellar sources to contribute more than half the total flux from the system \citep[e.g.][C10]{zurita02, orosz04}.
The ratio of nonstellar light to the total flux (hereafter denoted NSL fraction) is critical to measuring inclination accurately.
Several sources contribute to the NSL fraction, including the accretion disk, hotspots on the disk, and potentially a jet.
They can distort the shape of the photometric lightcurve in the following ways:
\begin{enumerate}
\item{The accretion disk contributes a baseline flux which dilutes the amplitude of the ellipsoidal variations.
To first order, increasing the disk flux is degenerate with lowering the inclination.}
\item{The disk and jet can exhibit aperiodic changes in brightness (flickering) due to the accretion flow (C10).
The timescale of this variability is short compared to the orbital period, so flickering can be misinterpreted 
as photometric error, particularly in folded lightcurves.  However, 
flickering is not a white noise process, nor is it stationary,
so binning and averaging the data will \textit{not} reproduce the underlying ellipsoidal shape.} 
\item{The disk hotspot can also distort the lightcurve shape.  The hotspot causes a peak in flux once per orbit,
which leads to assymmetric lightcurves when superposed
with double-peaked ellipsoidal variability.
Morever, the position of the hotspot is not constant, so it can alternately increase or decrease
the amplitude of the ellipsoidal variations for the same source (C10).}
\end{enumerate}  
Measuring inclination accurately is thus a delicate procedure that requires fitting
 a lightcurve with a consistent shape and a known NSL fraction, using a model that
includes a disk and a hot spot.
Using a less sophisticated model will result in systematically biased inclination measurements.
Moreover, because the lightcurve shape changes over time, different observing
runs will yield different inclination measurements, even if 
the same model is used to fit the lightcurve. 

\section{Quantification of Systematic Error for A0620-00}
We now characterize the systematic error in inclination measurements due to the effects discussed in the preceding section.
Our goal is to determine how inclination measurements obtained with a given model compare to the true inclination of a BHSXT.
Specifically, we want to know the distribution of inclination measurements one would obtain from observing a source at many distinct times, and 
where the true inclination lies within that distribution.
This requires two pieces of information: one, an accurate inclination measurement, and two, a description of the time variability of the source.

Such an analysis is possible for the source A0620-00, one of the best-studied BHSXTs.  Archival data spans the past 30 years, including eight observing runs and
more than a decade of daily photometric monitoring from the Small and Moderate Aperture Research Telescope System (SMARTS) consortium \citep{cantrell08}.
C10 use this extensive dataset to make an inclination measurement that accounts for the systematic errors discussed in $\S 2$.	 
They select 8 lightcurves that each maintain a consistent shape over the duration of the observations.
They fit every lightcurve with a 11-parameter model that includes an accretion disk with variable temperature, size and flaring angle and a hotspot with variable temperature,
size, and position.  During the fitting procedure, they constrain the disk to contribute the spectroscopically determined NSL fraction.
The best-fit inclinations to all 8 lightcurves are statistically self-consistent, giving a weighted average 
of $i = 51.0 \pm 0.9^\circ$.  We assume this value is unbiased by systematic error.  
	
Given an accurate inclination measurement for A0620-00, we now determine how inclination estimates vary with time.
First we select all subsets of existing data that resemble plausible observing runs.  We fit each of these
lightcurves with a star-only model to obtain a sample of inclination estimates. 
We chose a star-only model both because it is commonly used and because it introduces a straightforward systematic error. 

\subsection{Distribution of Inclination Measurements from Archival A0620-00 Data}
For our analysis, we use data collected by C10.
We select all subsets of this data that could be obtained on a typical observing run, i.e., many observations
in a short period of time.  Specifically, we require each subset to contain at least 20 data points over 7 days, with gaps in
phase no larger than 0.1.  We also specify that subsets do not overlap, ensuring that each point is only counted once.   
We restrict the sample to lightcurves of comparable quality to previously published BHSXT lightcurves by
binning the data in 30 phase bins and removing all lightcurves with average bin deviation  
greater than 0.03 mag.  Our final sample consists of 57 lightcurves taken with four different filters:  
V, I, H, and W (a wide bandpass centered at $4700 \mathrm{\AA}$).

We fit each lightcurve with a star-only model to obtain a distribution of inclination estimates.  
To fit the data, we use the Eclipsing Lightcurve Code (ELC) of \citet{orosz00}.	
Our model star has a mean temperature of 4600 K and a gravity darkening exponent of 0.10,
consistent with a K5 spectral type and a convective envelope \citep{lucy67, gray92}.
Limb darkening is computed directly from the model atmosphere.
For each lightcurve in our dataset, we obtain an inclination estimate $\hat{\imath}$ from the model that gives the lowest $\chi^2_\mathrm{red}$.
The distribution of $\hat{\imath}$ for all the lightcurves is shown in Figure 1.  
Note that we distinguish the estimated inclination $\hat{\imath}$ from the true inclination, denoted $i$.
We find $\hat{\imath}$ ranges from $33.9^{\circ}$ to $52.4^{\circ}$.  
Over $90\%$ of the measurements fall \textit{below} the C10 measurement of $\hat{\imath} = 51^{\circ}$.  
Such systematic underestimation is expected:  nonstellar flux almost always dilutes the amplitude
of the ellipsoidal variations, mimicking a lower inclination. 
The few lightcurves with $\hat{\imath} > 51^{\circ}$ 
may be the result of constructive interference between the hotspot and the ellipsoidal variability.
The distribution of $\hat{\imath}$ is bimodal, with one peak near $40^\circ$ and the other near $50^\circ$.  
We discuss potential mechanisms for this bimodality in the next section.

\begin{figure}
\begin{center}
\includegraphics[width = 0.45 \textwidth, angle=270]{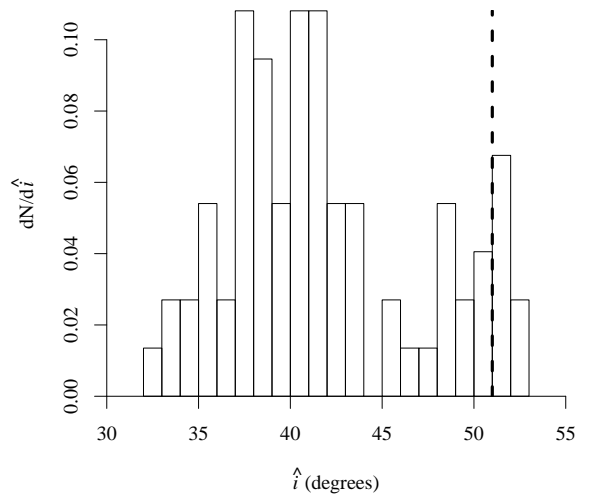}
\caption{Histogram of inclination estimates obtained by fitting a star-only model to extant A0620-00 lightcurves.  
Note that 90\% of the probability mass lies below the C10 inclination measurement $\hat{\imath} = 51^\circ$, marked in this plot with
a thick dashed line.}
\end{center}
\end{figure}

\subsection{Distribution of Inclinations from Simulated A0620-00 Lightcurves}
The bimodality in the distribution of inclinations seen in Figure 1 is an intriguing result
because it illustrates the potential impact of nonstellar flux on inclination measurements.
To explore how different factors contribute to the shape of the distribution, we introduce a method to simulate A0620-00 lightcurves.

We build on the work of \citet{cantrell08}, who analyze the time variability of A0620-00
using long-term photometric monitoring from the SMARTS consortium.  
They identify two distinct states --- labeled ``active" and ``passive" --- in the quiescent, optical A6020-00 lightcurve.
In the passive state there is minimal aperiodic variability, so the lightcurve shape is stable from night to night.  
By contrast, flickering is much more pronounced in the active state: active lightcurves are brighter, bluer, and more variable, 
possibly due to increased accretion activity. 
States persist for several months, but transitions from state to state occur on sub-night timescales.
A0620-00 is active roughly 70\% of the time (C10).
Because active and passive data behave differently, we simulate the two states separately.

\subsubsection{Simulation of Passive Data}
Although passive-state lightcurves do not change shape on short timescales, passive states separated by a period of activity 
\textit{do not} necessarily have the same lightcurve shape.
The 8 lightcurves fit by C10 were passive, but they had significantly different shapes from each other.
The main difference between the model fits were the hotspot parameters, suggesting that shape changes in passive
lightcurves are due to the changing temperature and position of the hotspot. 
Therefore, to model passive data, we select hotspot parameters 
from the range of fits in C10, listed in Table 1 (J. Orosz, private communication).
We find no correlation between any of the parameters, so we choose each parameter uniformly between the minimum and maximum 
value given in Table 1.
We use ELC to generate 500 passive V- and H- band lightcurves with the specified hotspot parameters.  We then add
photometric errors drawn from a normal distribution with mean 0 and standard deviation 0.03 mag.  
A representative sample of simulated lightcurves is shown in Figure 2.
The lightcurves are fit with a star-only model with parameters described in $\S 3.1$. 
All fits had $\chi^2_{\textrm{red}} < 3.0$.  We consider the quality of these fits acceptable,
as $\chi^2_\mathrm{red}$ has commonly exceeded 2.0 in previously published inclination fits
\citep[e.g.][]{shahbaz96, beekman97, vanderhooft98}. 
The distribution of best-fit inclinations for each filter is shown in the top panel of Figure 3.  
These distributions reveal that the A0620-00 hotspot
has a greater effect on the lightcurve shape at longer wavelengths.  For both filters, the distribution is 
peaked around $i = 51^{\circ}$ with some left skew.  
For an estimator $\hat{\imath}$ obtained 
by fitting a star-only model to passive data, the true inclination $i$ is 
given by $i = \hat{\imath}^{+3.3}_{-2.4}$ for H-band data and $i = \hat{\imath}^{+1.6}_{-1.1}$ for V-band.  
The limits denote 68\% confidence.  
These distributions reproduce the second mode in Figure 1, confirming the suggestion that the
hotspot can increase or decrease the amplitude of the ellipsoidal variations.
The distributions may be artificially narrow:  the range of
hotspot parameters listed in Table 1 is derived from only 8 fits, which probably do
not span the entire parameter space.  In addition to possible underestimation of the hotspot
variability, we neglect all other disk and jet variation.  
We therefore expect the distribution of $\hat{\imath}$ for passive lightcurves to 
be somewhat broader than the results quoted above.

\begin{figure}
\begin{center}
\includegraphics[width = 0.5 \textwidth]{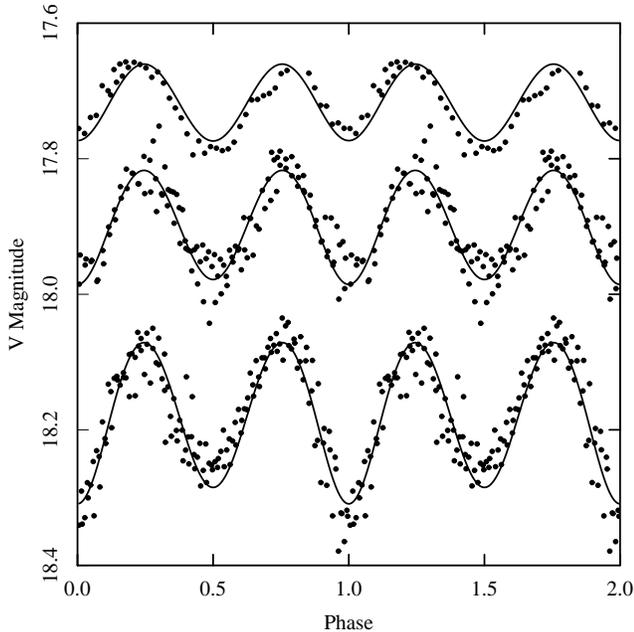}
\caption{Simulated active V-band lightcurves (points).  The best fitting star-only models (lines)
have inclinations $i = 30.4^\circ$, $37.4^\circ$, and $44.9^\circ$, with larger ellipsoidal variability corresponding to
higher inclination.  Each lightcurve consists of a single night of simulated data.}
\end{center}
\end{figure}

\begin{figure}
\begin{center}
\includegraphics[width=0.5 \textwidth]{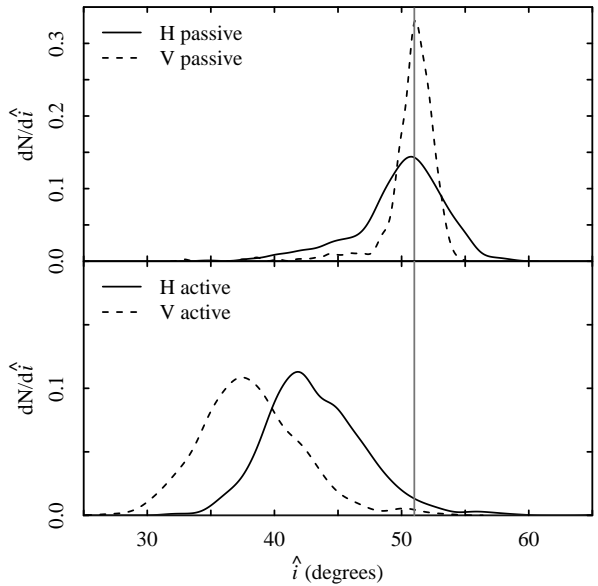}
\caption{Kernel density estimates for inclination obtained
from fitting a star-only model to passive and active lightcurves (top and bottom panels, respectively).  
Solid lines correspond to simulated H-band data and dashed lines to simulated V-band data.
All four samples contain 500 lightcurves.
The vertical gray line marks the inclination measurement obtained by C10.  }
\end{center}
\end{figure}

\ctable[
notespar,
nosuper,
mincapwidth = 0.47 \textwidth,
caption = Hotspot parameters,
width = 0.4\textwidth,
]{XXXXX}
{
	\tnote[\textbf{Notes:}]{This table lists the hotspot parameters for the ELC fits in C10.
	The ratio of hotspot temperature to disk temperature is $s_\mathrm{spot}$.
	The hotspot is centered at azimuthal angle $\theta_{\mathrm{spot}}$ and extends from the outer
	disk to an inner radius $r_{\mathrm{cut}}$.  The angular radius is given by $w_\mathrm{spot}$.
	}
}
{
\toprule
$\mathrm{Fit}$ & $s_{\mathrm{spot}}$ 	& $\theta_{\mathrm{spot}}$ & $r_{\mathrm{cut}}$ & $w_{\mathrm{spot}}$ \NN \midrule
1 	& 7.14		& 341.7 & 0.37 & 33.9 \NN 
2 	& 2.53		& 200.8 & 0.27 & 31.7 \NN 
3	& 1.53 		& 19.3 	& 0.87 & 90.0 \NN 
4 	& 2.41 		& 199.4 & 0.89 & 89.7 \NN 
5 	& 5.70 		& 190.3 & 0.07 & 70.7 \NN 
6 	& 4.20 		& 102.5 & 0.66 & 85.9 \NN 
7 	& 6.36 		& 72.5 	& 0.54 & 39.4 \NN 
8 	& 5.50 		& 37.7 	& 0.38 & 20.5 \NN \bottomrule
\addlinespace[15 pt]
}

\subsubsection{Simulation of Active Data}
For active-state data, we model the stellar and nonstellar components of the flux separately.
We simulate the stellar component with ELC using the same temperature and gravity darkening 
parameters as in $\S 3.1$.  We set the average stellar magnitude equal to the zero-disk magnitude found in C10.

Next we add a nonstellar component to the stellar lightcurve.
We model the nonstellar flux with a broken power law power spectral density (PSD).
The PSD cannot be determined directly from existing lightcurves due to severe daily and yearly aliasing.
We therefore develop an alternative method to characterize the time variability of the nonstellar flux.
We then determine the broken power law parameters that best reproduce that variability.

We begin by characterizing the time variability of the nonstellar flux.
Our data consist of H- and V-band active-state lightcurves from the Small and Moderate 
Aperture Research Telescope System (SMARTS) consortium, originally published in \citet{cantrell08}.  
The H-band dataset contains 1109 points over 2227 days and the V-band dataset contains
634 points over 2755 days.  
We subtract the ellipsoidal variability from this data using the stellar lightcurves
produced with ELC.  The median time between observations is one day;
however, there are large seasonal gaps in the data when the object is obscured by the sun.
Even when it is visible, sampling is somewhat sporadic.  Due to such uneven sampling,
conventional methods to characterize time variability fail.
We therefore use a modification of the binned autocorrelation function, denoted $C_\Delta(\tau)$, to describe the time variability.  

For the time series $\{(t_1,f_1),\ldots,(t_n, f_n)\}$:
\begin{equation}
C_{\Delta}(\tau)=\frac{1}{N} \sum_{i=1}^{n}
	(f_{i}-\bar{f}(t_i + \tau \pm \Delta))^2
\end{equation}

where 
\vspace{2mm}
\begin{compactitem}
\item{$\tau$ is the lag,}
\item{$\bar{f}(a \pm b)$ is the average of the set $\{f_{j}\}$ such that $\{t_j\}$ is within the range $a \pm b$}
\item{ and $N$ is the number of sets where $\{f_{j}\}\neq \varnothing $.}
\end{compactitem}
\vspace{2mm}
In other words, $C_\Delta(\tau)$ is the average squared difference for points separated by $\tau \pm \Delta$.
If there are no points separated by $\tau \pm \Delta$ (i.e., $N = 0$), $C_\Delta(\tau)$ is undefined.
To reduce the number of pairs that are likely to be correlated,
we do not include any $(t_i, f_i)$ more than once for a given lag $\tau$ .  For example, if $\tau = 10.0 \pm 0.05$ and
$t = \{0.0, 0.01, 10.0\}$, we remove the pair $\{0.01, 10.0\}$ because $t=10.0$ has already been included.
The error on $C_\Delta$ for a given $\tau$ is the sample standard deviation for the $N$ squared differences.
Note that these errors are not Gaussian, because pairs of points separated by $\tau \pm \Delta$ are not
statistically independent.

	The SMARTS observations provide sufficient coverage to constrain $C_\Delta(\tau)$ for lags on the order 
of $10^{-1}$ to $10^{3}$ days.  The median time difference between observations for both H- and V-band data is $1.0 \pm 0.05$ 
days, so we evaluate 
$C_\Delta(\tau)$ starting at $\tau = 1.0$ and increasing by factors of 2.0 to $\tau = 1024.0$.  
We choose $\Delta = 0.1 \textrm{ days}$.  
Variability on timescales $<0.1$ days is small compared
to variability for $\tau > 1.0$, so this choice of $\Delta$ does not bias the calculation of $C_\Delta$.
For $\tau \le 128.0$ days, this $\Delta$ gives $N > 300$ for both filters.  
On longer timescales, there are fewer pairs of points from
which to choose, so we exclude $\tau$ with $N < 15$.  
To constrain variability on short timescales, we also compute $C_\Delta(\tau = 0.1 \textrm{ days})$ with
$\Delta = 0.01$, obtaining $N = 25$ for the V-band data and $N=31$ for H.  
The results of the calculation of $C_\Delta$ are shown in Figure 4.

We next determine the broken power law PSD that best matches the $C_\Delta$ statistic for each data set.
We assume a PSD $S(f)$ of the form:
\begin{equation}
S(f) \propto
\left\{    , 
	\begin{array}{ll}
		\dfrac{1}{f^{\alpha}}  & \mbox{if } f \le f_b \\
		\dfrac{1}{f^{\beta - \alpha}_bf^{\beta}} & \mbox{if } f > f_b.
	\end{array}
\right.
\end{equation}

where $f$ is frequency, $f_b$ is the break frequency, $\alpha$ is the slope at frequencies less than $f_b$,
and $\beta$ is slope at frequencies greater than $f_b$.

To find the power law parameters that best reproduce the observed nonstellar lightcurve, 
we implement a grid-based $\chi^2$ minimization routine.  We loop
over $\alpha$, $\beta$, and $f_b$ and simulate a lightcurve from each set of parameters 
following the method of \citet{timmer95}.  We normalize the lightcurve so
that it has the same 25\% and 75\% quartiles as the observed nonstellar lightcurve.  
This normalization ensures that outlying data points do not bias the amplitude of the simulated lightcurve.
We then compute $C_\Delta(\tau)$ for the simulated data
and calculate its goodness of fit to $C_\Delta$ for the observed data.
We find the power law parameters that give the mininum $\chi^2$ are
$(\alpha, \beta, f_b) = (-0.8^{+0.1}_{-0.0},-1.5^{+0.7}_{-0.2},1.2^{+0.1}_{-0.7} \times 10^{-8} \mathrm{Hz})$ for V-band data, 
with $\chi^2_\mathrm{red} = 1.7$, and $(-0.8^{+0.7}_{-0.7}, -1.6^{+0.3}_{-0.4}, 5.0^{+8.0}_{-0.0} \times 10^{-9} \mathrm{Hz})$ for H-band, with $\chi^2_\mathrm{red} = 2.7$.
The results for the observed and best-fit simulated lightcurves 
are shown in Figure 4.
We obtain the uncertainties by holding two parameters fixed
and determining the range of the third parameter over which $\chi^2$ increases by a factor of two.
These ranges do not represent confidence intervals, nor do we expect $\chi^2_\mathrm{red} = 1$ for either data set,
because the errors on $C_\Delta(\tau)$ are not Gaussian. 
Nevertheless, varying the parameters within the quoted ranges does not significantly change the distribution
of $\hat{\imath}$ obtained from fitting the inclination of the simulated lightcurves.

\begin{figure}
\begin{center}
\includegraphics[width=0.5 \textwidth]{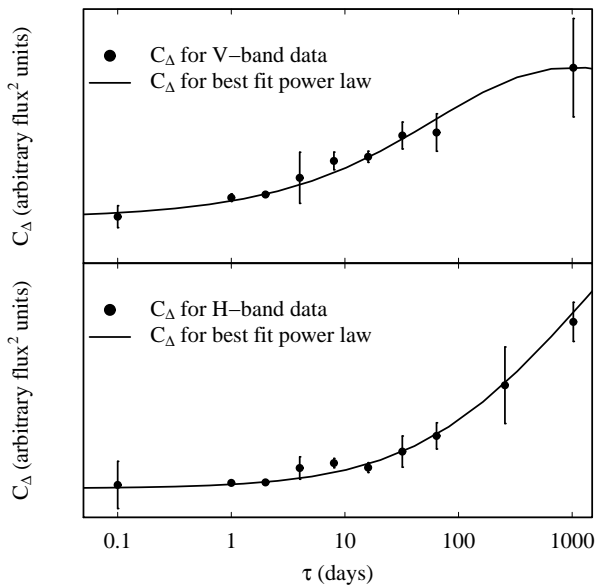}
\caption{The statistic $C_\Delta$ for V- and H-band SMARTS data (points in the top and bottom panel, respectively) and data 
generated from the best-fit power law (lines).  The value of $C_\Delta$ represents the average squared difference in flux between data points separated
by time $\tau \pm \Delta$. The error bars represent the standard deviation of $C_\Delta$ for the $N$ points associated with each $\tau$. 
The choice of $\tau$ for lags $>256.0$ is not identical for the two data sets because they are sampled slightly differently
and we require $N > 15$ for each $\tau$. The best-fit power law has parameters
$(\alpha, \beta, f_b) = (-0.8, -1.5, 1.2 \times 10^{-8} \mathrm{Hz})$ for V-band data and $(-0.8, -1.6, 5.0 \times 10^{-9} \mathrm{Hz})$ for H-band.}
\end{center}
\end{figure}

To obtain mock active-state data, we simulate fifteen 2500-day long lightcurves in V and H using the best-fit power law
parameters.  The lightcurves are sampled at 10-minute intervals.
We normalize each lightcurve such that the 25\% and 75\% flux quartiles match the observed SMARTS data.
We limit the length of the lightcurves to 2500 days because $C_\Delta(\tau)$ is poorly constrained for $\tau \gtrsim 1000$.
In addition, the normalization should be determined using lightcurves of comparable length.
After normalizing the nonstellar lightcurves, we add back the stellar component of the flux.

In order to select sample data from the 2500-day long simulated lightcurves, we recreate plausible observing runs in the following way:
first, we remove data points during ``daylight" hours, assuming 8 hours of viable observing time per 24-hour period.  We simulate
bad weather by removing 4-hours segments of data with probability 20\%.  Samples of the lightcurves are then selected with 
with mock observing runs separated by 30-day intervals.  Each run has equal probability of being one, two, or three consecutive days long. 
We exclude the run from the final sample if it has gaps in phase greater than 0.1.  We also bin each run in 30 phase bins
and restrict the sample to those runs with average bin standard deviation less than 0.03 mag and none greater than 0.1 mag,
as we did for subsets of real data.
We select 500 of the remaining lightcurves for each filter such that the samples are the same size. 
Several representative lightcurves are shown in Figure 2.  
We fit each of the lightcurves in the sample with the star-only model described in $\S 3.1$.  
The resulting estimates of inclination, again denoted $\hat{\imath}$, are shown in the bottom panel of Figure 3.  
The distributions are approximately normal.  We fit a Gaussian
to each distribution and find $\hat{\imath} \sim N(38.2^\circ,4.0^\circ)$
for the V-band fits and $\hat{\imath} \sim N(43.2^\circ,3.9^\circ)$ for H-band,
where the notation $N(\mu, \sigma)$ indicates a normal distribution with
mean $\mu$ and standard deviation $\sigma$.
The relation between true inclination $i$ and the estimator $\hat{\imath}$ is then given by 
$i \sim N(\hat{\imath} + 12.8^\circ, 4.0^\circ)$ for V and $i \sim N(\hat{\imath} + 7.8^\circ, 3.9^\circ)$ for H.

\subsection{Lessons from A0620-00}
Using simulated data, we find that fitting a star-only model to passive state lightcurves results in 
unbiased inclination estimates, approximately normally distributed around the true inclination.
The scatter can be explained by a hotspot with changing position and temperature.
On the other hand, fitting active data introduces
a bias towards artificially low inclination estimates.  The typical underestimation is
around $8^\circ$ for IR lightcurves and $13^\circ$ for optical lightcurves.  The source of the bias is a 
significant NSL fraction that increases with shorter wavelength.

These results reproduce the bimodality in inclination estimates obtained from past observations
(shown in Figure 1).  The optical state of the source is unknown for many of these observations, but 
the first mode near $\hat{\imath} \sim 40^\circ$ is comparable to the fits to active data and the second mode around
$\hat{\imath} \sim 50^\circ$ is consistent with the passive fits.
In addition, the relative weight of the two modes (roughly $2/3$ of the probability mass centered on the first mode)
matches the C10 observation that A0620-00 is active around $70\%$ of the time.

\section{Generalization of Systematic Effects to Other Systems}
We will now use the description of systematic effects for A0620-00 as a framework to estimate
the systematic error on inclination for other sources.  We treat passive and active data separately.

For passive A0620-00 data, we found that inclination estimates $\hat{\imath}$ obtained with a star-only model
are related to the true inclination $i$ by $i = \hat{\imath}^{+\sigma_1}_{-\sigma_2}$.
We assume a similar relation is valid for other systems, because true stellar ellipsoidal variability is dominant
in passive lightcurves.  Thus we expect inclination measurements made using passive data
to be unbiased (centered on the true inclination) for all systems. We choose $\sigma_1 = \sigma_2 = 3.0^\circ$.
The $\sigma$ values we obtained for A0620-00 are not symmetric and are somewhat smaller;
however, as discussed in $\S 3.2.1$, they are most likely underestimates, so $\sigma = 3.0^\circ$ is a 
conservative approximation of the error. 

For active A0620-00 data, we found
\begin{equation}
i \sim N(\hat{\imath} + \xi, \sigma).
\end{equation}
To generalize this result to other sytems, we scale the values of $\xi$ and $\sigma$ based on the system's
orbital parameters.  We find that $\xi$ can be computed from the NSL fraction, which we denote $\phi$.
We discuss this approximation in $\S 4.1$ and describe a method to estimate the NSL fraction 
in $\S 4.2$.  We chose $\sigma$ to scale linearly with $\xi$.

\subsection{Dependence of Inclination Measurements on the NSL Fraction}
We expect that the bias in inclination measurements for active data is determined primarily by the NSL fraction, and that
flickering and hotspots are secondary effects.  To demonstrate this, we simulate a set of lightcurves with two components:  a star and
a constant offset flux representing the NSL fraction.  We choose $i=51^\circ$ for the stellar lightcurve and NSL fractions ranging
from 0.0 to 0.9.
We fit these diluted lightcurves with a star-only model to obtain 
$\hat{\imath}$ as a function of the NSL fraction, $\phi$.  
A third-order polynomial fit to the results is
\begin{equation}
i - \hat{\imath} =  0.2 + 28.6 \phi - 15.3 \phi^2 + 27.0 \phi^3.
\end{equation}
where we take $i =51^\circ$.
We plot this fit in Figure 5 and overplot the points $(\phi, \hat{\imath})$ for each of the simulated A0620-00 active lightcurves.
These points fit the curve closely:
the simulated active data have $\hat{\imath}$ within $\pm 2.7^\circ$ of
the value obtained from adding a constant NSL fraction (limits denote 68\% confidence).
We therefore conclude that the NSL fraction is indeed the dominant factor determining the bias $\xi = i - \hat{\imath}$, with some scatter introduced by
the disk hotspot and flickering.
\begin{figure}
\begin{center}
\includegraphics[width = 0.5 \textwidth]{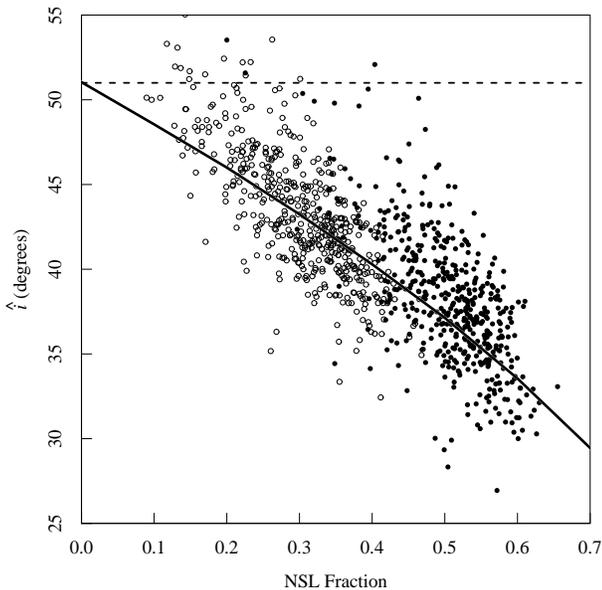}
\caption{Inclination estimates from a star-only model as a function of phase-averaged NSL fraction for
simulated active H- and V-band lightcurves (open and closed circles).
The line gives the inclinations fit to stellar lightcurves with a constant NSL fraction added. 
The active lightcurves have $\hat{\imath}$ within $\pm 2.7^\circ$ of the black curve at 68\% confidence.
The dashed line marks the inclination measurement of C10, $i = 51.0^\circ$.
}
\end{center}
\end{figure}

\subsection{Estimating the NSL Fraction Using Orbital Parameters}
Given that the NSL fraction determines the bias $\xi$ to first order, 
we would like to estimate the typical NSL fraction for sources other than A0620-00. 
Qualitatively, we expect a lower NSL fraction for systems with relatively hotter stars and
a higher NSL fraction for systems with relatively larger accretion disks.  To quantify 
these relationships, we focus on three observable parameters:
\begin{enumerate}
\item{Spectral type of the secondary star.  The hotter the star, the greater its total flux contribution
and the lower the NSL fraction.}
\item{Mass ratio.  The mass ratio determines the relative size of the Roche lobes in the system.
Assuming the fraction of the Roche lobe occupied by the disk is constant over all systems, a system 
with higher $q$ has a relatively larger disk and correspondingly higher $\phi$.} 
\item{Inclination.  As $i$ increases, the projected area of the disk decreases, thus lowering $\phi$.}  
\end{enumerate}

 To scale the NSL fraction based on the system parameters listed above, we use the Stefan-Boltzmann law.
For an optically thick source, the flux scales with the area of the emitting region and the temperature to the fourth power.
For convenience, we work with the ratio of stellar to nonstellar flux, which we denote $\rho$.
This parameter is related to the NSL fraction by $\rho = 1/\phi - 1$.
We expect $\rho$ to scale as:
\begin{eqnarray}
\rho 	&\propto& \left(\frac{T_\mathrm{star}}{T_\mathrm{disk}}\right)^4\left(\frac{R_\mathrm{star}}{R_\mathrm{disk}}\right)^2 \sec{i}  \\
	&\propto& \left(\frac{T_\mathrm{star}}{T_\mathrm{disk}}\right)^4\left(q^{0.45}\right)^2 \sec{i},
\end{eqnarray}
where $T$ and $R$ denote temperature and effective radius, with subscripts denoting the star and disk.  
Equation (7) replaces the ratio of radii with $q$ using an approximation from \citet{fkr02}.
The $\sec{i}$ factor accounts for the orientation of the accretion disk, assuming
it lies entirely in the orbital plane.  We approximate that the star is spherical, so its flux is independent of inclination. 


We can use the expression in equation (7) to estimate the NSL fraction of any source, provided we know the constant of proportionality.
To solve for the constant, we use the known parameters for A0620-00.  We use the estimates of $q$, $i$, and $T_\mathrm{star}$ from C10.
We estimate $\rho$ from the simulated active A0620-00 data, obtaining $\rho = 1.0 \pm 0.25$ in the optical and
$2.3 \pm 0.6$ in the IR.  The remaining unknown is $T_\mathrm{disk}$, which we choose to absorb into the constant of proportionality.
By neglecting the $T_\mathrm{disk}$ dependence, we effectively assume that
all sources have the same disk temperature as A0620-00.  This approximation is reasonable because the internal
dynamics of the disk determine its temperature, not the properties of the gas as it passes through the inner Lagrange point.  In addition, this 
assumption is substantiated by empirical evidence
from three cases:  the nonstellar light is redder than the B star in SAX J1819.3-2525 \citep{macdonald11}, bluer than the K star in 
A0620-00 (C10), and approximately constant across the optical spectrum for the F star in GRO J1655-40 (Cantrell, private communication).
Under this assumption, we can solve for the proportionality constant in (7) and 
thus infer $\rho\, \sec{i}$ for any system with known $q$ and $T_\mathrm{star}$.

We can use our estimate of $\rho \,\sec{i}$ to infer the bias $\xi$ in a star-only inclination estimate, $\hat{\imath}$.
First, we generate a set of stellar lightcurves and add a constant flux offset to each such that the ratio of stellar to nonstellar flux equals $\rho \,\sec{i}$.
We vary $i$ from $0^\circ$ to $90^\circ$ in $1^\circ$ intervals.  Second, we fit each lightcurve with a star-only model.  We then find the lightcurve whose inclination estimate
is closest to $\hat{\imath}$.  Because we know $i$ for this lightcurve, we can obtain the bias from $\xi = i - \hat{\imath}$ and the NSL fraction according to equation (5).  
To evaluate the accuracy of this procedure, we compute the NSL fraction from $\rho \sec{i}$ for all sources with star-only inclination estimates.  We compare these 
values to spectroscopically determined NSL fractions in Figure (6).  There is good agreement between our predictions and the observed values, with 7 of 11 predictions within $1\sigma$
from the observed value.  

The spread in the NSL fraction of A0620-00 introduces a small amount of scatter in our estimate of $\xi$, generally $< 2^\circ$.
This error is unrealistically narrow because it does not include any uncertainty due to other components of the lightcurve, such as flickering.
To estimate the $\sigma$ in equation (4), we choose the linear relation $\sigma = 3.0 + 0.115\xi$.
This scaling reduces to the expression for passive data in the case $\xi = 0.0^\circ$ and 
ensures that $\sigma = 3.9^\circ$ for $\xi = 7.8^\circ$ (i.e., the values we find for
simulated active H-band lightcurves).  

\subsection{Additional Constraints on Inclination}
There are two additional constraints on inclination we can obtain.  First, we set an upper limit based on the absence of eclipses.  
The visibility of eclipses depends on the size of the accretion disk and the mass ratio, $q$.  We assume a conservative
disk radius of $0.5 R_2$ and use ELC to determine the eclipse limit as a function of $q$.  
Systems with lower $q$ eclipse at higher inclinations,
so we use low-end estimates of $q$ to establish secure upper limits on $i$
for the all BHSXTs with measured mass ratios.  Specifically, we
use the minimum value for mass ratios described by a uniform distribution
and the value two standard deviations below the mean for mass ratios
with a normal distribution.  The upper limits calculated according to this procedure are
listed in Table 2.

The second constraint on inclination is a lower limit obtained by assuming the secondary star has
a mass equal to or less than that of a main sequence star of its spectral type. 
A Roche-lobe filling star is out of thermal equilbrium, so
the relationship between mass, radius, and surface temperature may be quite different from that of spherical stars.
Thus far, undermassive secondaries have been
observed in several sources, including GRO J1655-40 \citep{vanderhooft97} and A0620-00 (C10).
We therefore interpret the standard mass for a secondary star's spectral type as an upper limit to its true mass.
If $q$ is known and $m_*$ is overestimated,
$m_\mathrm{BH}$ is also overestimated.  By inspection of equation (1),
we find that overestimating $m_\mathrm{BH}$ results in an underestimation of $i$ for fixed $q$ and $f$.
There is some uncertainty in the spectral type for most of the sources in our sample, so
to obtain the most stringent lower limit, we calculate $i$ using the mass of the star with the earliest allowed spectral type,
along with the low-end estimates for $q$ and $f$ described in the preceding paragraph.  If the low-end estimate for $q$ is less than 0.0,
the lower limit on $i$ is also $0.0^\circ$.  The lower limits on inclination for all sources
are listed in Table 2.

\begin{figure}
\begin{center}
\includegraphics[width = 0.5 \textwidth]{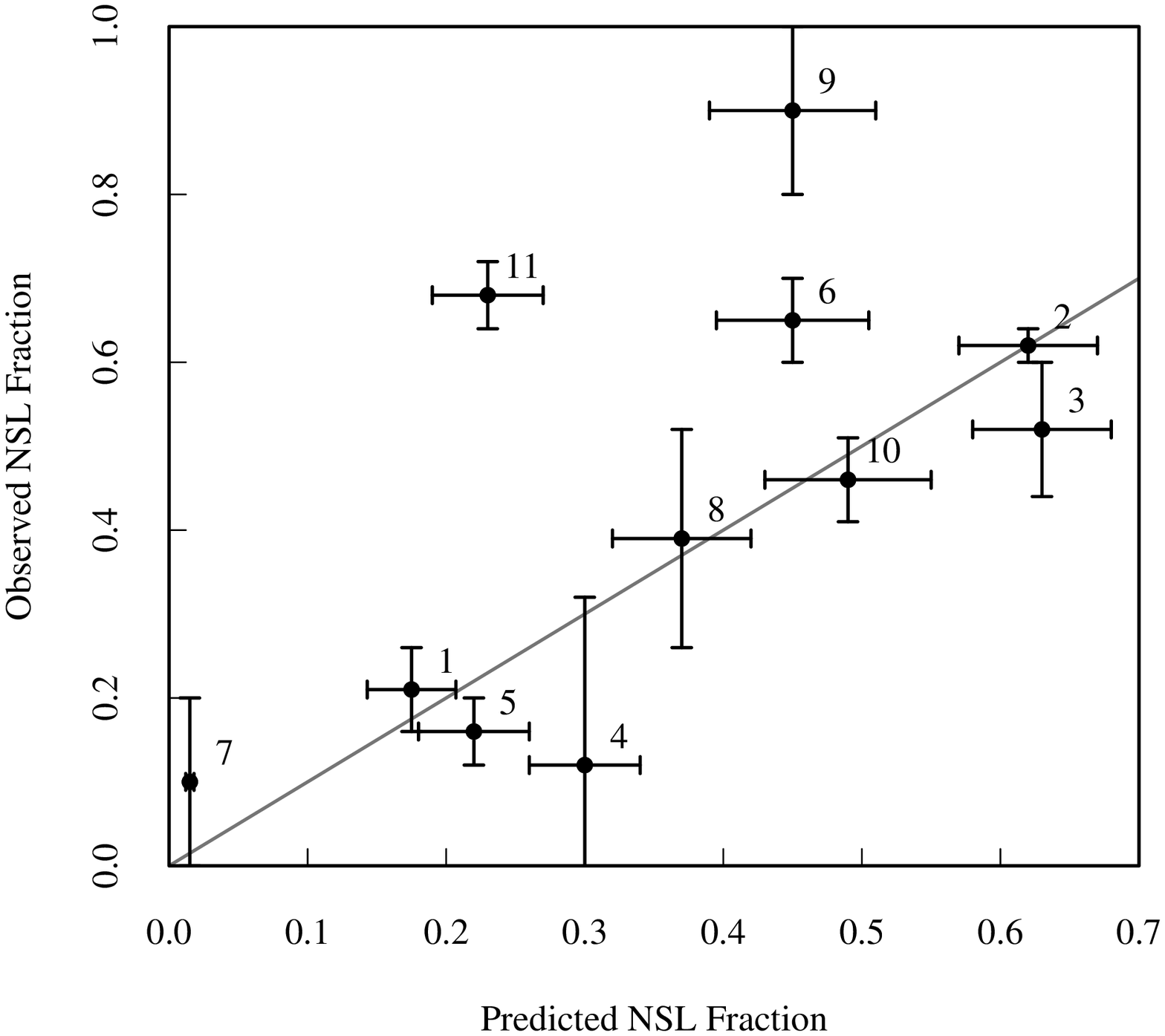}
\caption{Observed versus predicted optical NSL fractions (points).  The gray line is a visual aid respresenting perfect
agreement between the observations and predictions.  
The numbers indicate the reference for the observation:
(1) \citet{orosz98}, (2) \citet{webb00}, (3) \citet{casares95}, (4) \citet{shahbaz99},
(5) \citet{casares93}, (6) \citet{filippenko99}, (7) \citet{macdonald11}, (8) \citet{orosz02}, 
(9) \citet{orosz04}, (10) \citet{orosz96}, (11) \citet{wagner01}.
   The vertical error bars indicate either $1\sigma$ error or an allowed range. The horizontal error bars reflect the uncertainty due to
the variation in the NSL fraction of A0620-00, on which the predictions are based.  The excess in the observed NSL fraction for (9) may
because the source (GRO J1650-500) was not in full quiescence during the observation. 
The lack of agreement for source (11) (XTE J1118+480) may be due to this system's high inclination;
for large $i$, our assumption that the accretion disk lies entirely in the orbital plane may be invalid.}
\end{center}
\end{figure}

\section{Re-evaluation of Inclination and Mass Estimates for Transient Black Hole Binaries}

Using the procedure established in the previous section, we examine the inclination measurements for the black holes in our sample.
Much of the previous work on black hole X-ray binaries has had the goal of establishing a lower limit on the mass of the compact object to
demonstrate that it is a black hole and not a neutron star.  As a result, the conservative lower limits on mass are often secure,
but the masses and their errors may not be accurate. 
There a number of standard practices that may lead to erroneous measurements, which we briefly outline before discussing
individual objects in detail.  

One of the most frequent assumptions when fitting 
inclination is that the nonstellar flux is neglible, particularly in the infrared
 \citep{shahbaz94,shahbaz96,beekman97,gelino01, greene01, gelino03}.  
In many cases, this assumption has been justified by claiming the disk spectrum can be modeled as
a power law with negative slope, based on the precedent of \citet{oke77}.
The accretion disk thus contributes less total flux in the IR than in the optical.  
However, this approach has been called into question by recent measurements of 
significant IR NSL fractions in some sources \citep{reynolds08,gelino10}.  
In addition, there is evidence that the NSL fraction \textit{increases} with 
wavelength for some objects \citep{orosz97,reynolds07}.  
Inclinations obtained from fitting a star-only model should therefore be treated with caution.

Another common practice when fitting inclination is to combine lightcurves
from two or more observing runs \citep[viz.][]{shahbaz94,remillard96,gelino03,ioannou04,orosz04,gelino06,casares09,orosz11}.
However, due to aperiodic flickering and migratory hotspots, it is unlikely that 
combining and/or binning data reproduces the true underlying lightcurve shape. 
In addition, if the datasets being combined have non-overlapping phase coverage, 
large changes in the lightcurve shape may not be detectable.
Binning data that has significant intrinsic aperiodic variability tends to depress the amplitude of the ellipsoidal variations and thus lead
to systematically low inclination measurements.
 
A third potential source of error is using a NSL fraction determined at a different time from the photometric observations. 
The nonstellar flux can exhibit significant variability on timescales of months,
as we show in $\S 3.3$ for A0620-00.
 Similar variability has been observed in many other sources, including XTE J1550-564, SAX J1819.3-2525, 
GRO J0422+32, GS 1354-64, and XTE J1859+226 \citep{reynolds07, casares09,corralsantana11,macdonald11,orosz11}.
Given these changes in brightness, it is useful to determine the NSL fraction simultaneously 
with photometry to obtain an accurate inclination measurement. 

In the following section, we examine the data available for each of the 16 BHSXTs in our sample.  
In particular, we attempt to determine whether the source was observed in an active or passive state.
We then determine an appropriate value and error on the inclination, as described in $\S 4$.  In some cases,
this results in values of $i$ and hence of the black hole mass that are quite different from those in the literature.
More extensive observations will be required to check and refine these estimates.

\subsection{4U 1543-47}
4U 1543-47 is one of the few sources with near simultaneous photometry and spectroscopy.
\citet{orosz98} analyze B-, V-, and I-band photometry from 1998 June 28 - July 4 and
spectroscopy from 1998 July 1 - 4. 
The authors determine the NSL fraction in B, V, R, and I from the observed spectrum 
and use the measurements
as constraints when fitting models of ellipsoidal variability to the lightcurve.
Due to the relative brightness of the A-star secondary,
we expect the NSL fraction to be small.
A larger source of error is the mass ratio, which the
authors include as a free parameter in the model. 
They find $24^\circ < i < 36^\circ$ at the $3 \sigma$ level, 
but note the possibility of additional systematic effects.
We therefore use this range as the boundary of a uniform distribution.
A more precise measurement, $i \sim N(20.7^\circ, 1.5^\circ)$, was reported
in conference proceedings by \citet{orosz02}.  We do not use this measurement because
there is no published record of the lightcurve.  However, we note that using this smaller inclination would increase the most likely black
hole mass by more than a factor of two.

\subsection{A0620-00}
We adopt an inclination measurement of $51 \pm 0.9^\circ$ for A0620-00, based on the analysis of C10.
See $\S 3$ for a discussion. 

\subsection{GRO J1655-40} 
GRO J1655-40 is a well-studied source, with very regular quiescent lightcurves \citep{orosz97, greene01, beer02}. The consistency in
lightcurve shape exemplifies the connection between the stellar temperature and the NSL fraction that was discussed in $\S 3.3$:  GRO J1655-50 has
a bright F star secondary, so we expect that it has relatively small NSL fraction, and consequently, a stable lightcurve shape.   
Detailed studies of optical/IR emission and orbital parameters of GRO J1655-40 in full
quiescence have been published by \citet{orosz97}, \citet{greene01}, and \citet{beer02}.  The \citet{orosz97} study gives a tight constraint
on $i$ because their modeling requires a partial eclipse of the secondary to account for the depth of the minimum at phase 0.5.  By contrast, the \citet{greene01} analysis uses a more
sophisticated model for limb-darkening that allows for a deep primary minimum without eclipses.  The \citet{greene01} analysis 
use optical and IR lightcurves collected between 1999 July 7 and October 30.  
Over the course of observations, the lightcurves remained stable within the range of photometric error, a few hundredths of a magnitude.  The lightcurves are also consistent 
with those reported from several years before by \citet{orosz97}.  Thus these lightcurves appear to be passive.
The data are fit with a star-only model, giving a best-fit inclination $i = 70.2^\circ$ with
$\chi^2_\mathrm{red} = 1.6$.
\citet{beer02} obtain a similar result by reanalyzing the \citet{orosz97} data.  They use a model that includes an accretion disk and distance constraints obtained from the kinematics
of the radio jet, finding
a best fit inclination of $i = 69.0^\circ$ with $\chi^2_\mathrm{red} = 1.6$.  
Recent unpublished measurements of the NSL fraction (Cantrell, private communication) suggest that the distance to the source obtained by \citet{beer02} is more accurate than that of
\citet{greene01}.  We therefore adopt the \citet{beer02} measurement, which is compatible with the \citet{greene01} result, and assume the source was passive during the observations,
thus taking $i \sim N(69.0^\circ, 3.0^\circ)$.

\subsection{GRO J0422+32}
GRO J0422+32 has an unusually late-type secondary \citep{filippenko95}, and thus we expect the disk contribution to be especially strong. 
The inclination  has been frequently discussed
since it was first measured by \citet{orosz95}.  
That work finds $i > 45^\circ$ based on I-band photometry from
1994 October 27-28.  The lightcurve exhibits noticeable shape
changes between the two nights and a difference in mean I
magnitude of 0.05.  This type of variability is
typical of the active state.  A lower limit on $i$ is obtained by 
fitting a star-only model to ellipsoidal variability
with amplitude 0.15 mag, an approximate amplitude determined by considering each
night in the observed lightcurve separately. 
This measurement is consistent with the work of
\citet{gelino03}, who find $i = 45 \pm 2^\circ$ from fitting a star-only
model to J, H and K' lightcurves.  There is evidence, however,
for substantial IR disk contamination: \citet{reynolds07}
detect no ellipsoidal variability in H and K lightcurves that have
mean magnitudes consistent with the \citet{gelino03} lightcurves.
These conflicting results suggest that both the \citet{gelino03}
and the \citet{reynolds07} lightcurves are in the active state,
so we interpret the \citet{gelino03} measurement as a lower limit. 
 These results are consistent
with the work of \citet{filippenko95}, who find
$i = 48 \pm 3^{\circ}$, assuming a normal mass M2 V secondary 
and mass ratio $q = 0.1093\pm0.0086$. As discussed in $\S 4.3$, assuming a normal
mass secondary gives a lower limit on inclination.

In contrast to the above measurements, there are several 
works which find $i < 45^\circ$: \citet{casares95} obtain 
$i = 30 \pm 6^{\circ}$ for $q = 0.1$, using I-band data and assuming zero disk contribution.
However, there are significant gaps in phase coverage
near the primary maximum at phase 0.25.
Another low measurement is obtained by \citet{callanan96},
who find $i < 45^\circ$ for an assumed disk contamination less than 0.2.
This NSL fraction is based on what is considered typical for
other SXTs, which may not be applicable to this source because of the late-type secondary.
Moreover, the lightcurve they fit consists of 14 nights of data
between 1994 September 11 and 1995 January 9 that are combined
and binned.
Given the substantial aperiodic variability this source exhibits,
binning data over multiple nights may flatten
the lightcurve shape, implying a lower inclination.
Similarly, \citet{beekman97}, who find $10^\circ < i < 26^\circ$
use a binned R-band lightcurve obtained between 1995 January 12-15.
The lightcurve consists of just 55 points,
binned in phase bins of width 0.1.  

Given the large variability exhibited by GRO J0422+32, binning the lightcurves is unusually problematic.
In addition, those measurements fall well below the lower limit obtained using the spectral type of the secondary.
Therefore we discount measurements obtained by binning optical lightcurves. We adopt $\hat{\imath}  = 45^\circ$, assume the source is active,
and adjust the inclination according to $\S 4$, obtaining an inclination
$i \sim N(63.7^\circ, 5.2^\circ)$.  We note that the $3 \sigma$ lower limit
of this distribution is consistent with the lower limit of \citet{filippenko95}.  The peak of the probability distribution for the mass of the black hole
is $2.1 M_\odot$, which may call into the question the nature of this compact object.

\subsection{GRS 1009-45}
The only analysis of ellipsoidal variability in GRS 1009-45 was
performed by \citet{shahbaz96}.  They obtain R-band
photometry on 1995 May 8-10 and 1996 February 17-21.  The lightcurve
shows evidence of aperiodic variabity:  the peak brightness
fluctuates by around 0.1 mag during the time of observations, clear evidence of a significant nonstellar contribution.
\citet{shahbaz96} bin the data and fit a star-only model,
obtaining a best-fit inclination $i = 44^\circ$ with $\chi^2_\mathrm{red} = 3.2$.
Such a high $\chi^2$ indicates that
the lightcurve is most likely active, as is also suggested by the variability. 
We therefore adjust the inclination estimate $\hat{\imath} = 44^\circ$ 
according to $\S 4$, obtaining $i \sim N(62.0^\circ,5.1^\circ)$.

This result conflicts with the work of 
 \citet{filippenko99}, who determine inclination
assuming the secondary is not undermassive.
They find $i = 78^\circ$ for a K7-K8 star with mass
$m_* \approx 0.6 M_\sun$.  According to the 
reasoning in $\S 4.3$, this
inclination should be a \textit{lower} limit.  However,
there is some ambiguity about the spectral type.
\citet{dellavalle97} suggest the spectral type may be as early G5 V.  
Following the procedure outlined in $\S 4$, we find that a G5 V star implies a lower limit of
$\hat{\imath} = 42.3^\circ$, consistent with our estimate
$i \sim N(62.0^\circ,5.1^\circ)$.

\subsection{GRS 1124-68}
GRS 1124-68 is one of the few sources
for which simultaneous spectroscopy and photometry exist.
\citet{orosz96} analyze spectroscopy obtained 1992 April 3 and
photometry between 1992 April 3-15 in I and W (the W filter is a wide bandpass centered at $4700 \AA$).
Ellipsoidal modulation is easily discernible in this data.
By contrast, long-term monitoring of the source
with SMARTS shows no clear ellipsoidal variability
since 2003 (C. Bailyn, private communication).  
We therefore assume the source was passive during the 1992 observations 
and active post-2003.

Using the 1992 spectroscopy, \citet{orosz96}
constrain the B+V-band NSL fraction to $0.45 \pm 0.05$.
Because the source was passive during this period,
the NSL fraction is most likely valid for the 12 days
of photometry following the spectroscopic observations.
Using NSL fractions in the quoted range, \citet{orosz96} fit the 1992 B+V
lightcurve with a star + constant flux offset model.  However, the B+V lightcurve
has uneven maxima, a signature of the hotspot in the passive state.
Fits to the larger maximum yield inclinations above the eclipse limit,
so \citet{orosz96} fit the smaller maximum and obtain $54^\circ < i < 65^\circ$.

This measurement is somewhat higher and less precise than that of \citet{gelino01}.
They find $i = 54 \pm 1.5^\circ$ from a star-only fit to J and K 
lightcurves obtained 2000 February 20-21.  However, 
there is clear evidence for substantial nonstellar flux in the IR \citet{gelino10},
when the source was active.
The source may have been passive in 2000 February, 
but that does not guarantee a negligible NSL fraction. 

The \citet{orosz96} result is also slightly higher than that of 
\citet{shahbaz94}, who find $i = {54^{+20}_{-15}}^\circ$ for a 
folded, binned H-band lightcurve from 1995 February 17-20.
However, the best fit inclination gave $\chi^2_\mathrm{red} = 5.4$,
which the authors suggest is a consequence of 
incorrect sky subtraction and flat-fielding.  
We therefore adopt the \citet{orosz96}
result as the most reliable inclination measurement, $54^\circ < i < 65^\circ$.

\subsection{GRS 1915+105}
GRS 1915+105 is the only X-ray binary for which we can measure inclination
without analysing ellipsoidal variability.  
\citet{greiner01} determine an inclination $i = 70 \pm 2^\circ$ based
on the apparent superluminal motion of the jet.  They assume
the jet is perpendicular to the orbital plane due to a lack of observable
precession over several years of observation.
GRS 1915+105 has been active since its discovery in 1994 \citep{mirabelrodriguez94}.  
Single-hump variability and 
a superhump period have been detected \citet{bailyn11}, but
no studies of ellipsoidal variability have been possible to date
because of substantial disk and jet flux variation. 

\subsection{GS 1354-64}
\citet{casares09} obtain multiwavelength photometry and spectroscopy
between 1995 and 2003.  The data is characterized by strong aperiodic variability, and
no clear ellipsoidal modulation is discernable.
\citet{casares09} measure a NSL fraction of 0.67 in 2004
and 0.5 in 2006.  The average R-band magnitude varies by nearly 1 mag 
between 2002 and 2004.
Because of this strong variability, \citet{casares09} cannot determine a lower
limit to inclination based on ellipsoidal variations.  From the spectral type and eclipse limits, 
we find $27.2^\circ < i < 80.8^\circ$ according to the criteria in $\S 4.3$.
We assume an isotropic distribution $i \sim I(27.2^\circ,80.8^\circ)$, indicating that
$\cos{i}$ is uniform between $\cos{27.2^\circ}$ and $\cos{80.8^\circ}$.

\subsection{GS 2000+25}
The most extensive study of ellipsoidal variability
in GS 2000+25 was performed by \citet{ioannou04}.  
They obtain I-band data from 1992 August and 1998 September,
as well as R-band data from 1998 September-October, 1999 June, July, and September, and
2000 July-August.  The authors note some changes in lightcurve shape between the 
observing runs:  for example, the R-band lightcurve has equal
maxima in 2000 August, but unequal maxima in 1998 September.  These changes
are small, however, and there is relatively little scatter in the folded lightcurve.
We therefore conclude that the source is passive during the \citet{ioannou04} observations.
The authors bin the lightcurve and fit a star-only model, finding $54^\circ < i < 60^\circ$.
Fits including a disk and hotspot give lower $\chi^2_\mathrm{red}$ values, but do not
affect the allowed range of inclinations.  Fits to data between phase 0.0 and 0.5,
which the authors suggest are less affected by the hotspot,
also produce the same result.  This lack of sensitivity to changes in the model
may well be a consequence of combining and binning data taken over many nights:
features such as a hotspot are obscured because the hotspot position
changes over time.  We suspect that binning the data depresses the \citet{ioannou04} inclination measurement:
even for passive lightcurves, binning slightly decreases the amplitude of the ellipsoidal variations.
Indeed, the \citet{ioannou04} measurement is somewhat lower than that of \citet{callanan96}, who
find $55^\circ < i < 65^\circ$ for a star-only fit to J and K' photometry obtained on 1995 August 11-12,
assuming $q < 0.05$.  The \citet{callanan96} result is most likely passive, because it 
gave a higher inclination range than the passive \citet{ioannou04} lightcurves.
Both the \citet{ioannou04} and \citet{callanan96} results are consistent with the \citet{beekman96}
estimate $43^\circ < i < 60^\circ$. 
For our final estimate, we adopt the \citet{callanan96} result
because it is less affected by binning than that of \citet{ioannou04},
given that the observations span only two days.
We take $i \sim I(55.0^\circ,65.0^\circ)$ as our inclination estimate. 

\subsection{GS 2023+338}
Strong aperiodic variability in the lightcurve of GS 2023+388 has
thus far prevented precise inclination measurements.
\citet{wagner92} obtained I-band photometry on 17 nights between
1990 September and 1992 May. During this period, the average I magnitude
varied by more than 0.1 mag, so the authors are only able to constrain $50^\circ < i < 80^\circ$.  
The lower limit is derived based on the observation of double-peaked Balmer lines in the spectrum, and
the upper limit is based on the lack of eclipses.
\citet{shahbaz94} find a similar allowed range, $45^\circ < \hat{\imath} < 83^\circ$, using K and K' photometry
obtained between 1992 August and 1993 December.  Their best fit with a star-only model has $\hat{\imath} = 56.0$ with a $\chi^2_\mathrm{red} = 15.2$.
They suggest the poorness of the fit may be due to an incorrect color correction between the K and K' data.
We therefore regard the limits obtained by \citet{wagner92} as more secure.

\citet{sanwal96} also fit a star-only model to IR data:  they obtain
H-band photometry on 17 nights between 1993 June and November.
Strong aperiodic variability is present in the lightcurve.  The
authors note that there is significantly more scatter in the observations of GS 2023+388
than in the lightcurve of the comparison star.  They also note a systematic increase in brightness
over one 6-hour period of observation.  At the beginning of that night, the data is fainter than the ellipsoidal model, but the source 
smoothly increases in brightness over 6 hours until it is nearly 0.1 mag brighter than the model lightcurve. 
This hour-scale time variability suggests that the source was active during these observations.
Such behavior is typical of the active state.  On the other hand, \citet{khargharia10} find almost
no disk contamination in the IR spectrum.  However, these data were obtained in 2007, and thus do not
speak to the state of the system when the \citet{sanwal96} data were taken over a decade earlier.
We therefore use the minimum inclination estimate of \citet{sanwal96}, $\hat{\imath} > 62.0^\circ$,
which they obtain using a star-only model with standard limb and gravity darkening.  We assume that the source was active at the time.  Using the method described in $\S 4$, we 
obtain $i \sim N(80.1^\circ, 5.1^\circ)$.  
We note that this is approximately equal to the eclipse limit.

\subsection{H1705-250}
The inclination of H1705-250 was first measured by \citet{martin95}.
They obtain 87 R-band images on 1992 May 1-6 and fit a star-only
model, finding $48^\circ < i < 51^\circ$.  The best fit had
$\chi^2_\mathrm{red} = 0.87$. 
The authors only show a folded lightcurve,
so it is difficult to detect aperiodic variability. 

\citet{remillard95} obtain a conflicting result using B+V observations from 1992 May, 1993 April, and 1994 July. 
The lightcurve exhibits uneven maxima, which suggests the source was active during the observations.  However,
the authors suggest this may be an artifact of changes in phase coverage between the observations, 
They find a lower limit on $i$ by fitting a star-only model to the lightcurve, excluding the higher maximum
from the fit.  Their fit restricts $i > 60^\circ$. 

Given the uncertainty in whether the source was active or passive for both these analyses,
we adopt the \citet{martin95} $i = 48^\circ$ estimate as a lower limit.
We cannot obtain an upper limit due to eclipses for this source because the limits on the mass ratio extend to 0.0.
Our final estimate for the inclination is thus $i \sim I(48.0^\circ, 90.0^\circ)$.

\subsection{SAX J1819.3-2525}
\citet{orosz01} analyze the photographic B-band lightcurve obtained by \citet{goranskij90}.
The photometric errors are large (near 0.1 mag), so the binned lightcurve
is imprecise.  The best-fit model to the lightcurve has $i = 70^\circ$
and a partial eclipse of the secondary by a large, faint accretion disk; however,
it systematically underestimates the amplitude of the ellipsoidal variability.

\citet{macdonald11} have compiled 10 years of data on this source, which they separate into clearly 
defined passive and active states.  Analysis of the passive data shows no nonstellar contribution, a result
that is supported in some cases by simultaneous spectroscopy.  However, the amplitude of the ellipsoidal
variability requires an extreme inclination of $\sim 90^\circ$.  The X-ray data are sparse enough that
an eclipse geometry cannot be excluded.  Therefore, we adopt $i \sim I(80.0^\circ, 90.0^\circ)$.

\subsection{XTE J1118+480}
There is consensus in the literature that XTE J1118+480 has a high inclination,
in the range $68^\circ < i < 81^\circ$.  However, there are a number of factors that
make an accurate measurement challenging.  One is a strong superhump
modulation in addition to the ellipsoidal variability \citep{zurita02}.
Another complicating aspect is the large and variable NSL fraction:
\citet{wagner01} measure an R-band NSL fraction of 0.72 for
2000 November 30 which decreases to 0.64 by 2001 January 4. 

\citet{gelino06} obtain data in B, V, R, J, H, and K that may be passive.
Their R-band lightcurve is around 0.8 mag fainter than that of \citet{wagner01},
and there is no detectable superhump period, both indications of less
disk activity.  They fit data in all bands
simultaneously and constrain the NSL fraction in each band.
They find $i = 68 \pm 2^\circ$. 
However, the authors assume there is zero disk contamination in the
H-band when fitting the spectral template,
an assumption that may be problematic.

\citet{wagner01} measure $i = 81\pm2^\circ$ with a star+disk model using a \textit{brighter} R-band lightcurve
than that of \citet{gelino06}.  This is unusual, because lightcurves with very high
levels of disk contamination generally cause
lower inclination measurements.  However, these data
were taken on three separate occasions
between 2000 December 14 and 2001 January 9, during which time the measured R-band disk
fraction varied significantly.  Moreover, the lightcurve consists of 68 exposures 
that were binned in 25 phase bins.  The combination of binning small amounts of data
taken weeks apart during a period of intense disk activity may lead to an unrepresentative
lightcurve shape.  

\citet{mcclintock01} also measure a very high inclination, finding a best fit $i = 80^\circ$ for a disk
fraction of 0.66.  However, they acknowledge uncertainty in the NSL fraction and set a lower limit
$i > 40^\circ$ for zero disk light.

The most extensive study of ellipsoidal variability in XTE J1118+480 is that of
\citet{zurita02}.  They obtain R-band photometry on 53 unique nights
between 2000 December and 2001 June.  They observe a steady decrease in
magnitude over this time interval, so they detrend the lightcurve by
subtracting the average nightly flux.  This lightcurve shows
evidence of superhump modulation and may be distorted by flickering
on sub-orbital timescales.  They infer a NSL fraction for the detrended lightcurve by extrapolating
the 2001 April measurement of \citet{wagner01} and assuming the decline in brightness
is due to a decrease in disk flux.  Using the extrapolated NSL fraction, they find an inclination in the range 
$71^\circ < i < 82^\circ$.  

None of the inclination measurements discussed above are free of significant
systematic sources of error; they are, however, reasonably consistent.
We adopt the full range of inclination estimates obtained for this object, taking $i \sim I(68.0^\circ, 82.0^\circ)$.

\subsection{XTE J1550-564}
\citet{orosz11} determine the inclination of XTE J1550-564 using photometry and spectroscopy obtained between
2001 and 2008.  The photometric observations include optical data from 2001 June and NIR
data taken between 2006 and 2008.  
The measured V- and R-band NSL fractions are 
0.3 and 0.39 using 2001 and 2008 spectroscopy, respectively.
There are definite changes in the NIR lightcurve shape between 2006-2007 and 2008,
so it it unlikely that the 2008 NSL fraction measurement is valid
for the 2006-2007 observations.  It is also unlikely that the 2001
NSL fraction is valid for any of the NIR observations.
\citet{orosz11} acknowledge this uncertainty, but fit the data using 
eight different combinations of lightcurves and NSL fractions.
There are four data subsets:  optical data only, optical data and 2006-2007 NIR data,
optical data and 2008 NIR data, and all optical and NIR data.  Each subset
is fit separately for both NSL fraction measurements.
The model includes a disk with four free parameters: radius, flaring angle, inner radius temperature, and temperature profile.
The lowest $\chi^2_\mathrm{red}$ is obtained using the optical lightcurves from 2001 and the 2008 NSL fraction.
However, as discussed in $\S 3.3$, using NSL fractions determined at a different time from
the observations produces unreliable inclination measurements. 
Nevertheless, the range of inclination measurements from all eight combinations is reasonably narrow:
$57.7^\circ < i < 77.1^\circ$.  Because we do not know which of the combinations is most appropriate, we 
adopt $i \sim I(57.7^\circ, 77.1^\circ)$ as our estimate of inclination.
We followed the same reasoning to obtain an estimate for the mass ratio, assuming $q$ is uniformly distributed
between the minimum and maximum values found by \citet{orosz11}.

\subsection{XTE J1650-500}
\citet{orosz04} determine the inclination of J1650-500 with R-band photometry
obtained between 2003 May and August.
The authors set a lower limit on inclination of $i > 50^\circ$
by fitting a star-only model. 
\citet{orosz04} also attempt to constrain the NSL fraction using spectroscopy from 2002 June,
tentatively finding disk contamination near 0.8 in the R-band.  There is no indication,
however, that this measurement is applicable to the 2003 photometry.
No $\chi^2_\mathrm{red}$ is given for the star-only fit, but we infer that the source
was active for the 2003 observations because
there appears to be more scatter in the folded lightcurve than one expects due to photometric error.  
We therefore assume the source was active and scale the lower limit $\hat{\imath} = 50^\circ$ 
according to $\S 4$, which implies $i \sim N(75.2^\circ, 5.9^\circ)$.

\subsection{J1859+226}
\citet{corralsantana11} obtain R-band photometry on 2008 July 31 - August 1
and 2010 July 13-14.  They compare this data to the 2002 R-band lightcurve
of \citet{zurita02} and find an increase in brightness of $\sim 0.25$ from
2002 to 2008 and $\sim 1.0$ from 2002 to 2010.  No clear ellipsoidal
modulation is detectable in the 2010 data, 
but the 2000 and 2008 data show variability with amplitude  0.3-0.4 mag, consistent with the passive state.
\citet{corralsantana11} find that a star-only model with $i = 60^\circ$
reproduces these data.  
We therefore adopt this value as representative of the passive state, 
and estimate $i = 60 \pm 3.0^\circ$.

\ctable
[
caption = Orbital parameters for 16 black hole binaries,
doinside = \footnotesize,
star,
nosuper]
{ccccccccc}
{
\tnote[\textbf{Notes:}]{This table gives the mass function, mass ratio, spectral type, and inclination estimate for 16
black hole binaries.  The measurements of $f$, $q$, and spectral type are taken from the literature.  
The lower limits on inclination, $i_\textrm{min}$, are
obtained assuming the secondary has a normal mass for its spectral type.  The upper limits, $i_\textrm{max}$, 
are the highest inclinations that do not cause eclipses.  For a detailed explanation of these limits and the
inclination estimates, see $\S 4$.
}
\tnote[\textbf{Notation:}]{
The notation $N(\mu, \sigma)$ implies a normal distribution with mean $\mu$ and standard deviation
$\sigma$.  A uniform distribution from $\alpha$ to $\beta$ is indicated by $U(\alpha, \beta)$.  An isotropic distribution
is denoted $i \sim I(\alpha, \beta)$, implying $\cos{i}$ is uniform between $\alpha$ and $\beta$. 
} 
\tnote[\textbf{References:}]{
	(1) \citet{orosz03}; (2) \citet{orosz98}; (3) \citet{neilsen08}; (4) \citet{cantrell10}; 
	(5) \citet{shahbaz99}; (6) \citet{greene01}; (7) \citet{webb00}; (8) \citet{harlaftis99}; (9) \citet{filippenko99};
	(10) \citet{shahbaz96}; (11) \citet{dellavalle97}; (12) \citet{orosz96}; (13) \citet{casares97}; (14) \citet{greiner01}; 
	(15) \citet{harlaftis04}; (16) \citet{casares09};
	(17) \citet{casares04}; (18) \citet{harlaftis96}; (19) \citet{casares95}; (20) \citet{casares94}; (21) \citet{filippenko97};
	(22) \citet{harlaftis97}; (23) \citet{orosz01}; (24) \citet{gonzalezhernandez08}; (25) \citet{calvelo09}; 
	(26) \citet{orosz11}; (27) \citet{orosz04}; (28) \citet{corralsantana11}
	}
}
{
\hline \addlinespace[0.8mm]
X-ray Name 	& Optical 	& Spectral Type		& $f (M_\sun)$		& $q$	 		& 	$i$	& $i_{\textrm{min}}$	& $i_{\textrm{max}}$	& refs. \NN 
		& Counterpart	&			&			&			&   (deg.)	& (deg.)		& (deg.)		& \NN  \addlinespace[0.8mm] \hline \addlinespace[0.8mm]
4U 1543-47 	& IL Lup 	& A2 V			& $N(0.25, 0.01)$ 	& $U(0.25, 0.31)$	& $I(24.0,36.0)$&	19.8		&	73.4		& 1,2 \NN \addlinespace[0.5mm]
A0620-00 	& V616 Mon 	& K5 V			& $N(3.1, 0.04)$ 	& $N(0.060, 0.004)$ 	& $N(51.0,0.9)$	&  	36.5		&	79.8		& 3,4 \NN \addlinespace[0.5mm]
GRO J0422+32 	& V518 Per	& M2 +2/-1 V		& $N(1.19, 0.02)$ 	& $N(0.116, 0.08)$	& $N(63.7,5.2)$	&	0.0		&	90.0		& 7,8 \NN\addlinespace[0.5mm]
GRO J1655-40 	& V1033 Sco	& F6 III		& $N(2.73, 0.09)$ 	& $N(0.38,0.05)$ 	& $N(69.0,3.0)$	&	0.0		&	90.0		& 5,6 \NN\addlinespace[0.5mm]
GRS 1009-45 	& MM Vel	& G5-K7 V		& $N(3.17, 0.12)$ 	& $N(0.137, 0.015)$ 	& $N(62.0,5.1)$ &	42.3		&	76.9		& 9,10,11 \NN\addlinespace[0.5mm]
GRS 1124-683	& GU Mus	& K3-K4 V		& $N(3.01, 0.15)$	& $N(0.128, 0.04)$	& $I(54.0,65.0)$&	33.2		&	80.1		& 12,13	\NN\addlinespace[0.5mm]
GRS 1915+105 	& V1487 Aql 	& K0-7 III		& $N(9.5, 3.0)$ 	& $N(0.058, 0.033)$ 	& $N(70.0,2.0)$ &	0.0		&	90.0		& 14,15 \NN\addlinespace[0.5mm]
GS 1354-64 	& BW Cir 	& G0-5 III		& $N(5.73, 0.29)$ 	& $N(0.12, 0.04)$ 	& $I(27.2,80.8)$&	27.2		&	80.8		& 16,17 \NN\addlinespace[0.5mm]
GS 2000+25 	& QZ Vul	& K3-6 V		& $N(5.01, 0.12)$ 	& $N(0.042, 0.012)$ 	& $I(55.0,65.0)$&	28.3		&	86.7		& 18,19 \NN\addlinespace[0.5mm]
GS 2023+338 	& V404 Cyg 	& K0 IV			& $N(6.08, 0.06)$ 	& $N(0.060, 0.005)$	& $N(80.1,5.1)$ &	35.4		&	80.0		& 20 \NN\addlinespace[0.5mm]
H1705-250 	& V2107 Oph	& K5$\pm$2 V		& $N(4.86, 0.13)$	& $U(0, 0.053)$ 	& $I(48.0,90.0)$&	0.0		&	90.0		& 21,22 \NN\addlinespace[0.5mm]
SAX J1819.3-2525& V4641 Sag 	& B9 III		& $N(2.74, 0.12)$ 	& $N(0.67, 0.04)$ 	& $I(80.0,90.0)$ &	44.8		&	69.6		& 23 \NN\addlinespace[0.5mm]
XTE J1118+480 	& KV UMa 	& K5 V			& $N(6.27, 0.04)$ 	& $N(0.024, 0.009)$ 	& $I(68.0,82.0)$ &	21.8		&	89.4		& 24,25 \NN\addlinespace[0.5mm]
XTE J1550-564 	& V381 Nor 	& K3$\pm$1 III		& $N(7.65, 0.38)$ 	& $U(0.031,0.037)$ 	& $I(57.7,77.1)$&	26.5		&	82.0		& 26 \NN\addlinespace[0.5mm]
XTE J1650-500	& \nodata	& G5-K4 III		& $N(2.73, 0.56)$	& $U(0,0.5)$		& $N(75.2, 5.9)$&	0.0		&	90.0		& 27 \NN\addlinespace[0.5mm]
XTE J1859+226 	& V406 Vul 	& K5-7 V		& $N(4.5, 0.6)$ 	& $U(0,0.5)$ 		& $N(60.0,3.0)$ &	0.0		&	90.0		& 28 \\ \addlinespace[0.5mm]
\hline \addlinespace
}

\section{The Black Hole Mass Distribution}
\label{sec:mass-dist}

We now analyze the mass distribution of black holes using the orbital
parameters and new inclination estimates listed in Table 2.  Previous
analyses of the mass distribution indicate that there is a ``mass
gap,'' or dearth of black holes, between the maximium theoretical
neutron star mass ($\approx 3 M_\odot$) and the mimimum black hole
mass \citep{bailyn98,ozel10,farr10}.  The presence of a mass gap has
important implications for the physics of black hole formation, as discussed
by \citet{belczynski11}.

Based on our arguments in $\S 5$ that many of these published masses
may be overestimates, it is plausible that the mass gap inferred in
earlier work is the result of systematic errors in mass measurements.
Using the revised system parameters in Table 2, we show in \S \ref{sec:mass-gap} that
there is no evidence of a mass gap, even when using models of the
black hole mass distribution that give strong evidence for a gap in
previous analyses.  We demonstrate in \S \ref{sec:mass-gap} that this
conclusion rests on the properties of one system in our sample,
GRO J0422+32.  However, it is not clear that the inclination correction we have
identified by analogy to A0620-00 is entirely appropriate for this system; we
discuss this issue in \S \ref{sec:0422-inclination}.  Further
observations of the system will be needed to settle this question.

\subsection{Statistical Methods}

To address the impact of systematic error in the mass distribution, we
repeat a subset of the Bayesian analysis in \citet{farr10} using the
adjusted inclination measurements discussed in $\S 5$.  Using the
distributions of system parameters from Table 2, we can compute the
probability distribution for the true mass of each system.  Figure 7
shows these distributions for all 16 BHSXTs.

\begin{figure*}
\begin{center}
\includegraphics[width=\textwidth]{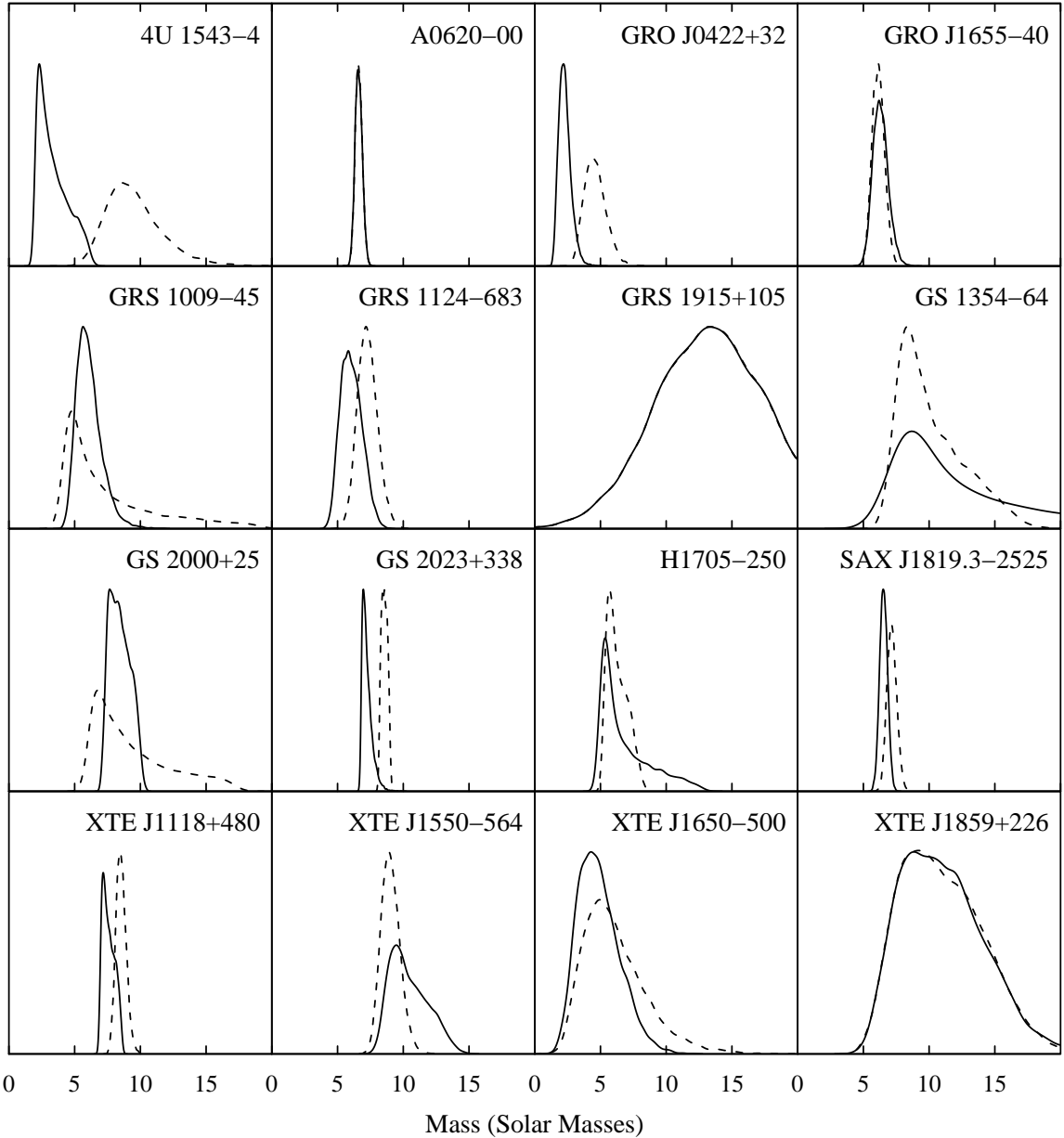}
\caption{\label{fig:masses} Kernel density estimates of the probability distributions for black hole
 mass implied by the system parameters in Table
 2 (solid lines).  The distributions implied by the system parameters used in \citet{farr10} are also shown
(dashed lines).  Each panel is normalized so that the probability distributions integrate to 1.}
\end{center}
\end{figure*}

For a set of parameters, $\theta$, governing the underlying BH mass
distribution and observational data, $D$, Bayes' rule states
\begin{equation}
\label{eq:Bayes}
P\left(\theta\mid
D\right)=\frac{P\left(D\mid\theta\right)P\left(\theta\right)}{P\left(D\right)}.
\end{equation}
Here $P\left(\theta\mid D\right)$, called the \emph{posterior}, is the
probability of obtaining particular values for $\theta$ given the
observational data, $D$. $P\left(D\mid\theta\right)$ called the
\emph{likelihood}, is the probability of obtaining the observed data,
$D$, when the underlying mass distribution is described by parameters
$\theta$. $P\left(\theta\right)$, called the \emph{prior}, is the
probability distribution one would expect for the parameters before
seeing the data. $P\left(D\right)$, called the \emph{evidence}, is a
normalization constant that ensures that the posterior is a proper
probability distribution on parameter space.

Our data $D$ is the set of probability distributions for the
underlying masses derived from the observations discussed in \S 5 and
displayed in Figure \ref{fig:masses}.  Write $P_i(m_\mathrm{BH})$ for
the distribution of masses in system $i$.  We assume that the mass
measurements for the 16 systems are independent; then the likelihood
is given by
\begin{equation}
P(D\mid\theta)=\prod_{i}\int
P(m_\mathrm{BH}\mid\theta)P_i(m_\mathrm{BH})dm_\mathrm{BH}.
\end{equation}
It is possible to evaluate the posterior in Equation \eqref{eq:Bayes}
directly for various values of the parameters, $\theta$.  However, this calculation is
computationally expensive in multi-dimensional parameter space.
As a more efficient alternative, we draw parameter \emph{samples} from the posterior
distribution via Markov chain Monte Carlo (MCMC) methods 
\citep{metropolis53,farr10}.
Given a set of samples, $\left\{ \theta_i
\right\}$, one can compute probability distributions for individual
parameters by histogram, and can approximate posterior-weighted
integrals over parameter space by sums,
\begin{equation}
 \left\langle f(\theta)\right\rangle_{p(\theta\mid D)} = \int d\theta
 \, f(\theta) p(\theta\mid D) \approx \frac{1}{N} \sum_i f(\theta_i).
\end{equation}

\subsection{Model}

We focus on the power-law model from \citet[Equation (7)]{farr10}
because it was the most-favored model for the LMXB mass distribution
out of the 10 considered in that work%
\footnote{We have verified that the qualitative behavior of the mass
 gap described in this section is present for the other models of the
 LMXB mass distribution considered in \citet{farr10}.}. %
The power law model has
\begin{multline}
 \label{eq:power-law}
 P\left(m_\mathrm{BH} | \left\{ M_\mathrm{min}, M_\mathrm{max},
    \alpha \right\}\right) = \\
 \begin{cases}
  A m_\mathrm{BH}^\alpha & M_\mathrm{min} \leq m_\mathrm{BH} \leq
  M_\mathrm{max} \\
  0 & \mathrm{otherwise}
 \end{cases},
\end{multline}
with parameters
\begin{equation}
 \theta = \left\{ M_\mathrm{min}, M_\mathrm{max}, \alpha \right\}.
\end{equation}
The normalization constant, $A$, is
\begin{equation}
 A = \frac{1+\alpha}{M_\mathrm{max}^{1+\alpha} - M_\mathrm{min}^{1+\alpha}}.
\end{equation}
We use uniform priors on $M_\mathrm{min}$, $M_\mathrm{max} >
M_\mathrm{min}$, and $\alpha$ within broad ranges that allow for black
hole masses up to 40 $M_\odot$:
\begin{multline}
 P(\theta) = \\
 \begin{cases}
  2 \frac{1}{40^2} \frac{1}{28} & 0 \leq M_\mathrm{min} \leq
  M_\mathrm{max} \leq 40, -15 \leq \alpha \leq 13 \\
  0 & \mathrm{otherwise}
 \end{cases}.
\end{multline}

In the analysis of \citet{farr10}, the power-law model had strong
evidence of a mass gap between the theoretical maximum mass of the
heaviest neutron stars ($\sim 3M_\odot$) and the mass of the lightest
black holes, defined as the 1\% quantile ($M_{1\%}$) of the mass
distribution in Equation \eqref{eq:power-law}.  \citet{farr10} had
$M_{1\%} > 4.3 M_\odot$ with 90\% confidence for the power-law model.

\subsection{The Mass Gap}
\label{sec:mass-gap}

\begin{figure}
 \begin{center}
  \includegraphics[width=\columnwidth]{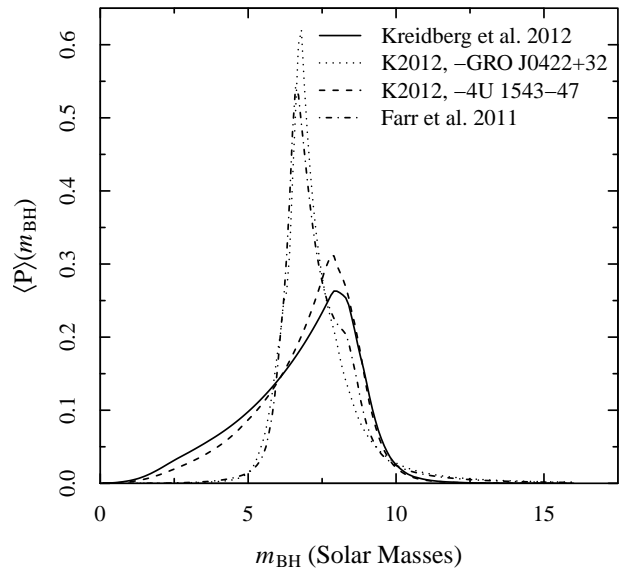}
 \end{center}
 \caption{\label{fig:power-law-dist} The marginalized mass
  distribution, Equation \eqref{eq:marginalized-pM}, for the
  power-law mass distribution model (Equation \eqref{eq:power-law}).
	The solid curve from an analysis using the
  system parameters in Table 2; the dotted curve from an analysis with
  the parameters from Table 2, but \emph{excluding} GRO J0422+32 from
  the sample; the dashed curve comes from an analysis
  with system parameters from Table 2, but \emph{excluding} 4U 1543-47 from the sample; 
  and the dot-dashed curve comes from the analysis using the system parameters
  of \citet{farr10}; The analysis using parameters from Table 2 shows
  significant support in the gap region, but this conclusion depends
  strongly on the updated parameters for GRO J0422+32.}
\end{figure}

With the revised system parameters in Table 2, there are two systems
whose mass distribution peaks below $4 M_\odot$: 4U 1543-47 and
GRO J0422+32.  Thus it is not surprising that the expected mass
distribution under the power-law model, defined by
\begin{equation}
 \label{eq:marginalized-pM}
 \left< P \right> (m_\mathrm{BH}) \equiv \int d\theta P(m_\mathrm{BH}
 \mid \theta) P(\theta \mid D)
\end{equation}
gains support in the mass gap.  That is, the existence of two sources whose
mass distribution is strongly peaked below $4M_{\odot }$ necessarily
requires that the overall mass distribution not be zero in that region.  
In Figure \ref{fig:power-law-dist}, we
show the $\left< P \right>(m_\mathrm{BH})$ implied by the
\citet{farr10} analysis and the $\left< P \right> (m_\mathrm{BH})$
implied by the system parameters in Table 2.  Interestingly, the extra
support in the gap region is due almost completely to the shift in the
mass distribution of GRO J0422+32; in Figure \ref{fig:power-law-dist} we
also show $\left< P \right>(m_\mathrm{BH})$ with parameters from Table
2, but \emph{excluding} GRO J0422+32 from the sample, and similarly for
excluding 4U 1543-47.  When 4U 1543-47 is excluded from the analysis,
the curve is essentially the same as when the entire set of 16 systems
is analyzed, while excluding GRO J0422+32 produces a curve that is very
close to that of \citet{farr10}.

\begin{figure}
 \begin{center}
  \includegraphics[width=\columnwidth]{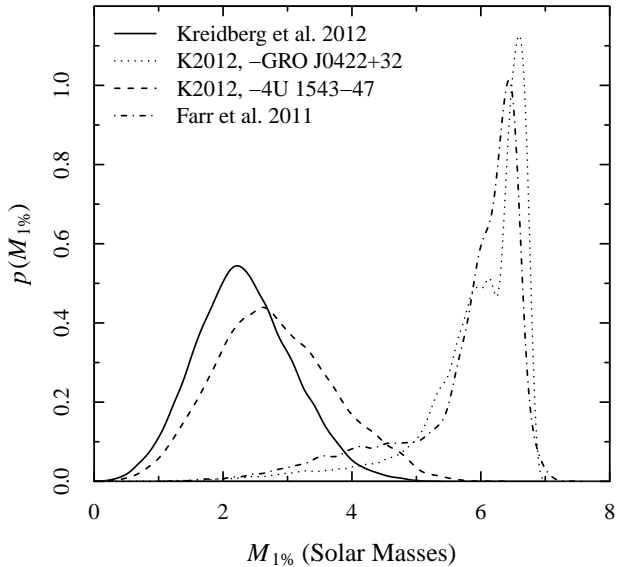}
 \end{center}
 \caption{\label{fig:power-law-mmin} The distribution of the 1\% mass
  quantile, $M_{1\%}$, in the power-law model implied by the
  analysis with system parameters from Table 2 (solid line);
  parameters from Table 2, but
  excluding GRO J0422+32 (dotted line);  parameters from Table 2, but
  excluding 4U 1543-47 (dashed line); and parameters from
  \citet{farr10} (dot-dashed line).  For the complete set of parameters
  from Table 2, and this set excluding 4U 1543-47, the range of
  likely minimum black hole masses extends through the gap ($M_{1\%}
  \lesssim 4 M_\odot$); when GRO J0422+32 is excluded the minimum black
  hole mass is equivalent to the analysis from \citet{farr10}. }
\end{figure}

Another way to address the presence of the mass gap is to examine the
posterior probability distribution of the 1\% mass quantile,
$M_{1\%}$.  Figure \ref{fig:power-law-mmin} shows this distribution
from the \citet{farr10} parameter values, the updated parameters in
Table 2, and using the updated parameters but excluding 4U 1543-47 or
GRO J0422+32.  The analysis using the complete set of systems from Table 2
has significant probability for $M_{1\%} \lesssim 3 M_\odot$; so does
the analysis with parameters from Table 2, but excluding 4U 1543-47.
On the other hand, the analysis with parameters from \citet{farr10}
and Table 2, but excluding GRO J0422+32, have $M_{1\%} \gtrsim 4.3
M_\odot$ with $90\%$ confidence.  Thus, the presence or absence of a
mass gap depends strongly on the properties of the mass distribution
for GRO J0422+32.  In addition, if in the future the mass distribution of 4U
1543-47 were constrained to a narrow peak near the low-mass end
of its current distribution, this system would also provide strong
evidence against a mass gap.  Because the continued evidence for or
against the mass gap may depend on the constraints on these systems
from future observations, we next discuss the current status of
observations of these systems in more detail.

\subsubsection{GRO J0422+32} \label{sec:0422-inclination} 
\begin{figure}
 \begin{center}
  \includegraphics[width=\columnwidth]{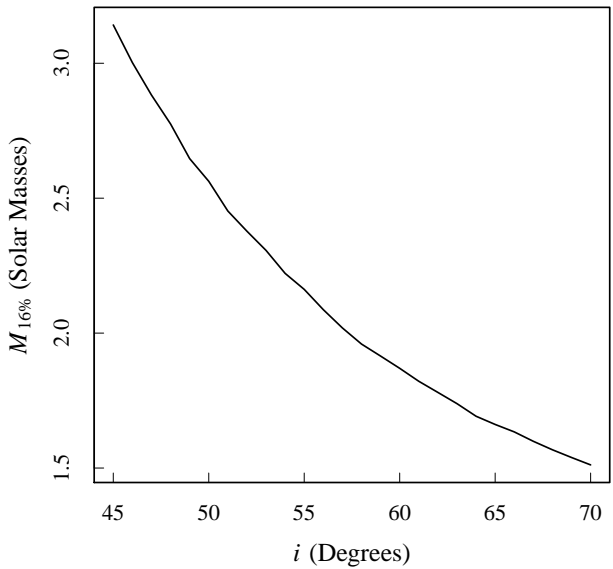}
 \end{center}
 \caption{\label{fig:j0422-inc} The variation in the 16\% quantile of
  J0422+32's mass distribution, $m_{16\%}$, with the assumed peak of
  the inclination distribution for this object.  $m_{16\%}$ is a
  rough proxy for the peak of the $M_{1\%}$ distribution (see Figure
  \ref{fig:power-law-mmin}).  As long as the inclination of J0422+32
  is peaks at $i \gtrsim 45^\circ$ we expect to obtain no evidence
  for a mass gap.}
\end{figure}

The
inclination of GRO J0422+32 hinges on whether this source is generally
observed in an active state, and whether that active state biases the
inclination measurements in a similar way to that of A0620-00.  As
shown above, if the typical measurements of $i \sim 45^\circ$ are accepted,
then the previously identified mass gap still exists.  However, if it is
larger, as is suggested by
the analogy with A0620-00, then the compact object in this system must
be roughly between 2 and 3 solar masses, which effectively fills the
mass gap (see the relevant panel in Figure 7).  In Figure \ref{fig:j0422-inc} we show how the 16\% quantile
of J0422+32's mass distribution varies as its inclination
distribution's mean varies from the previously-accepted value to the
value in Table 2, holding the width of the inclination distribution
fixed.  The 16\% quantile approximately tracks the peak of the
$M_{1\%}$ distribution shown in Figure \ref{fig:power-law-mmin}, so we
can see the effect on the mass gap from varying the inclination
without repeating the analysis for each value of $i$.  We expect to
obtain no evidence of a mass gap as long as the inclination
distribution of J0422+32 peaks at $i \gtrsim 45^\circ$.

In many ways, GRO J0422+32 is just the kind of system one would expect
to be heavily biased by non-stellar light.  The secondary star is one
of the smallest and coolest of the entire sample.  Therefore changes
in the non-stellar flux will be reflected strongly in the overall flux
of the system.  And indeed, GRO J0422+32 is quite faint in quiescence
($R\sim 21$) and strongly varying lightcurves are observed at
different times \citep[compare][]{curry03,gelino03}.  Thus
it seems likely that the observed lightcurves of this source are
distorted by non-stellar light in ways that change strongly with time.
Whether these distortions conspire to create significant
underestimates in the inclination, as in A0620-00, is not yet clear.
One route to resolving this issue would be to carry out simultaneous
photometry and spectroscopy of GRO J0422+32, so that the division of stellar and
non-stellar light can be determined for each individual observation,
resulting in an unbiased light curve of stellar light only.  However,
given the faintness of this source, such observations will be
challenging.

\subsubsection{4U1543-47} 
The other source which has a mass probability distribution significantly
lower than in previous work is 4U1543-47.  This is because we did not
accept the \citet{orosz02} inclination value, since no published lightcurve
accompanied that report, but rather accepted the earlier, significantly 
higher estimate of Orosz et al. (1998).  That previous higher inclination measurement
resulted in a significantly smaller black hole mass, but still allowed masses
above $4M_{\odot }$, so the source does not greatly affect the possibility
of a mass gap.  However, if the observations become more precise at the higher
inclination value this source might also be an issue for the mass gap, so we
discuss it further here.

The difficulties with 4U1543-47 are
associated primarily with its low inclination.  The $\sin^3 i$ term in 
equation (1) means that at low inclination very small changes
in $i$ create large changes in the black hole mass.  Going from the
\citet{orosz02} result of $i\simeq 20^\circ$ to the previous result centered
around $i \simeq 30^\circ$ results in a change in the centroid of the probability
distribution of the black hole mass of more than a factor of two.  At
low inclinations, the amplitude of the ellipsoidal modulation is also
quite small ($\lesssim 0.1$ magnitudes in this case), so high precision
magnitude measurements are required.  Given that errors of a few
degrees in inclination determinations seem inevitable, it may prove
difficult to pin down the inclination of 4U 1543-47 sufficiently
tightly to fully resolve the issue.

4U 1543-47 does have some observational advantages, however.  It is one of
the three sources (with SAX 1819.3-2525 and GRO J1655-40) that have
early-type secondaries.  These systems all have very regular
ellipsoidal variations in the passive state, with very little
non-stellar light; this is presumably because the relatively hot
secondary stars are particularly bright.  Like the other early
secondary systems, 4U1543-47 is quite bright in quiescence ($V\sim
15$) so that also makes the system easy to study.  There is one
technical difficulty in studying 4U1543-47: it is in the wings of an
even brighter star, but this can be dealt with through appropriate use
of PSF fitting techniques.  A long term photometric data set for this
source, similar to those for A0620-00 (C10) and SAX1819.3-2525
\citep{macdonald11} has been obtained, and is currently being
analyzed.

\subsection{Discussion}
\label{subsec:discussion}
A resolution of the specific situation of GRO J0422+32 and 4U1543-47 will be
required to determine whether a true mass gap exists in the BHSXT mass distribution.
However, even if these two objects do prove to have $M<4M_{\odot }$, it is
still true that the mass distribution rises from the low mass end.  Thus the relative
paucity of low mass black holes in these systems still requires an explanation.
The peak of the mass distribution appears to be around
$8M_{\odot }$, slightly higher than in the 
previous analysis by Farr et al. (2011). The new analysis also confirms another
aspect of the previous analysis, namely a sharp cutoff above $10M_{\odot }$.
This is quite different for the BHSXT black hole distribution than for the
wind-fed systems with $M_\mathrm{star}>M_\mathrm{BH}$ (e.g. Cyg X-1 and M33 X-7), which often
contain much more massive black holes.  The difference presumably arises
due to the dramatically different evolutionary scenarios that create these
two kinds of systems \citep{valsecchi10, wong12}. It is important to avoid combining these two kinds of
systems into a single mass distribution, since there are clearly significant
empirical and evolutionary differences between them.

The characteristics of the BHSXT mass distribution should provide
guidance and constraints for the physics of massive core collapse.
Previous studies have
discussed possibilities for observational biases
and/or physical mechanisms that can explain the characteristics of the
low-mass BH distribution in these systems
\citep{brown01,fryer01,ozel10}. Most recently, \citet{belczynski11}
have carefully assessed how this low-mass tail and the potential
presence of a gap leads to constraints regarding the timescale for the
instability growth that eventually drives the stellar explosion, as
well as the role of binary evolution in affected the BH masses in
low-mass X-ray transients. Our current results, with
the mass gap partially filled in, still provide an empirical basis 
for exploring the physical processes considered in these studies,
although the constraints on the explosion timescale will be weaker.

\section{Conclusions}
We have explored the systematic effects of non-stellar light on the
mass estimates of black holes in BHSXTs.  We summarize our conclusions
as follows:

\begin{enumerate}
\item{By examining the case of A0620-00, for which the most extensive and
carefully analyzed data set exists, we find that observations in the
``active" state significantly underestimate the inclination of the system,
and consequently overestimate the black hole mass.
By contrast, observations in the ``passive" state appear to be relatively
unbiased toward the system inclination.}
\item{We estimate how large such effects might be in other
systems, based on the assumption that the accretion flows are roughly
similar in form, while the temperature of the star and the relative
size of the star and the accretion disk vary from system to system.  The assumption of similar
accretion flows appears to be in good agreement with spectroscopic measurements of the NSL fractions.}
\item{We re-examine the literature on the sixteen known black holes in SXTs.
The data and analysis of many of these systems has been aimed at establishing
a firm minimum for the mass of the compact object, since that is what
establishes the identity of the compact object as a black hole.  When examined
from the point of view of determining accurate inclinations, and thus
accurate black hole masses, we identify a number of problematic issues.

These problems stem from the presence of variable non-stellar light, even in the IR.  
The amount of this
extra flux can vary significantly from one observation run to another, and thus
it is not safe to use a measurement of stellar light fraction from one time to calibrate
photometric observations made at another time.  The non-stellar light also
appears to vary with orbital phase, so the assumption that it is a flat ``dilution''
of the ellipsoidal variations is not always true. Significant variability also means
that phase binning over long periods of time may result in a lightcurve that is
unrepresentative of the underlying ellipsoidal variations.}
\item{When we apply a consistent set of criteria to the existing literature on
black hole SXTs, we find that two objects, GRO J0422+32 and 4U1543-47, may
have small black hole masses (below 4-5 $M_\odot$, see Figure 7).  Specifically, the low mass required for GRO J0422+32
eliminates the mass gap identified
in previous work on the black hole distribution.   But this
result depends on the active state of GRO J0422+32 biasing the inclination measurement
in the same way as in A0620-00.  While this seems plausible,
additional observational work will be required to
assess whether the active state is indeed biasing the black hole mass in this system.  
We note that if GRO J0422+32 is excluded from the analysis presented here, the results from
previous studies are reproduced quite closely.}
\item{However, even if GRO J0422+32 (and possibly 4U1543-47) do fall
into the purported mass gap, the basic features of the mass distribution of
BHSXTs remain: there are relatively few low mass ($<5M_{\odot }$) black holes;
there is a peak in the distribution around $7-8M_{\odot }$; and a sharp dropoff
in numbers beyond $10M_{\odot }$.  This distribution is quite different from the 
``high-mass'' black hole binary systems, and provides interesting constraints
on the supernovae and binary evolution processes that create BHSXTs.}

\end{enumerate}

We are grateful for illuminating discussions with Andrew Cantrell in the early stages of this work.
We thank Ritaban Chatterjee and Jerry Orosz for assistance with ELC and helpful suggestions.  
We appreciate useful feedback from Rachel MacDonald and Ilya Mandel.
LK acknowledges support from a Yale Science Scholars fellowship.
CB and LK acknowledge support from NSF grant AST-0707627.
WMF acknowledges support by CIERA (Center for Interdisciplinary Exploration and Research in Astrophysics); WMF and VK acknowledge partial support from NSF grant AST-0908930.

\bibliography{mybib}{}

\begin{thebibliography}{80}
\expandafter\ifx\csname natexlab\endcsname\relax\def\natexlab#1{#1}\fi

\bibitem[{{Bailyn} \& {Buxton}(2011)}]{bailyn11}
{Bailyn}, C.~D., \& {Buxton}, M. 2011, in American Astronomical Society Meeting
  Abstracts \#218, \#229.05--+

\bibitem[{{Bailyn} {et~al.}(1998){Bailyn}, {Jain}, {Coppi}, \&
  {Orosz}}]{bailyn98}
{Bailyn}, C.~D., {Jain}, R.~K., {Coppi}, P., \& {Orosz}, J.~A. 1998, \apj, 499,
  367

\bibitem[{{Beekman} {et~al.}(1996){Beekman}, {Shahbaz}, {Naylor}, \&
  {Charles}}]{beekman96}
{Beekman}, G., {Shahbaz}, T., {Naylor}, T., \& {Charles}, P.~A. 1996, \mnras,
  281, L1

\bibitem[{{Beekman} {et~al.}(1997){Beekman}, {Shahbaz}, {Naylor}, {Charles},
  {Wagner}, \& {Martini}}]{beekman97}
{Beekman}, G., {Shahbaz}, T., {Naylor}, T., {Charles}, P.~A., {Wagner}, R.~M.,
  \& {Martini}, P. 1997, \mnras, 290, 303

\bibitem[{{Beer} \& {Podsiadlowski}(2002)}]{beer02}
{Beer}, M.~E., \& {Podsiadlowski}, P. 2002, \mnras, 331, 351

\bibitem[{{Belczynski} {et~al.}(2011){Belczynski}, {Wiktorowicz}, {Fryer},
  {Holz}, \& {Kalogera}}]{belczynski11}
{Belczynski}, K., {Wiktorowicz}, G., {Fryer}, C., {Holz}, D., \& {Kalogera}, V.
  2011, ArXiv e-prints

\bibitem[{{Brown} {et~al.}(2001){Brown}, {Heger}, {Langer}, {Lee}, {Wellstein},
  \& {Bethe}}]{brown01}
{Brown}, G.~E., {Heger}, A., {Langer}, N., {Lee}, C.-H., {Wellstein}, S., \&
  {Bethe}, H.~A. 2001, NewA, 6, 457

\bibitem[{{Callanan} {et~al.}(1996){Callanan}, {Garcia}, {Filippenko},
  {McLean}, \& {Teplitz}}]{callanan96}
{Callanan}, P.~J., {Garcia}, M.~R., {Filippenko}, A.~V., {McLean}, I., \&
  {Teplitz}, H. 1996, \apjl, 470, L57+

\bibitem[{{Calvelo} {et~al.}(2009){Calvelo}, {Vrtilek}, {Steeghs}, {Torres},
  {Neilsen}, {Filippenko}, \& {Gonz{\'a}lez Hern{\'a}ndez}}]{calvelo09}
{Calvelo}, D.~E., {Vrtilek}, S.~D., {Steeghs}, D., {Torres}, M.~A.~P.,
  {Neilsen}, J., {Filippenko}, A.~V., \& {Gonz{\'a}lez Hern{\'a}ndez}, J.~I.
  2009, \mnras, 399, 539

\bibitem[{{Cantrell} {et~al.}(2008){Cantrell}, {Bailyn}, {McClintock}, \&
  {Orosz}}]{cantrell08}
{Cantrell}, A.~G., {Bailyn}, C.~D., {McClintock}, J.~E., \& {Orosz}, J.~A.
  2008, \apjl, 673, L159

\bibitem[{{Cantrell} {et~al.}(2010){Cantrell}, {Bailyn}, {Orosz}, {McClintock},
  {Remillard}, {Froning}, {Neilsen}, {Gelino}, \& {Gou}}]{cantrell10}
{Cantrell}, A.~G., {et~al.} 2010, \apj, 710, 1127

\bibitem[{{Casares} \& {Charles}(1994)}]{casares94}
{Casares}, J., \& {Charles}, P.~A. 1994, \mnras, 271, L5

\bibitem[{{Casares} {et~al.}(1993){Casares}, {Charles}, {Naylor}, \&
  {Pavlenko}}]{casares93}
{Casares}, J., {Charles}, P.~A., {Naylor}, T., \& {Pavlenko}, E.~P. 1993,
  \mnras, 265, 834

\bibitem[{{Casares} {et~al.}(1995){Casares}, {Martin}, {Charles}, {Martin},
  {Rebolo}, {Harlaftis}, \& {Castro-Tirado}}]{casares95}
{Casares}, J., {Martin}, A.~C., {Charles}, P.~A., {Martin}, E.~L., {Rebolo},
  R., {Harlaftis}, E.~T., \& {Castro-Tirado}, A.~J. 1995, \mnras, 276, L35

\bibitem[{{Casares} {et~al.}(1997){Casares}, {Martin}, {Charles}, {Molaro}, \&
  {Rebolo}}]{casares97}
{Casares}, J., {Martin}, E.~L., {Charles}, P.~A., {Molaro}, P., \& {Rebolo}, R.
  1997, NewA, 1, 299

\bibitem[{{Casares} {et~al.}(2004){Casares}, {Zurita}, {Shahbaz}, {Charles}, \&
  {Fender}}]{casares04}
{Casares}, J., {Zurita}, C., {Shahbaz}, T., {Charles}, P.~A., \& {Fender},
  R.~P. 2004, \apjl, 613, L133

\bibitem[{{Casares} {et~al.}(2009){Casares}, {Orosz}, {Zurita}, {Shahbaz},
  {Corral-Santana}, {McClintock}, {Garcia}, {Mart{\'{\i}}nez-Pais}, {Charles},
  {Fender}, \& {Remillard}}]{casares09}
{Casares}, J., {et~al.} 2009, \apjs, 181, 238

\bibitem[{{Chen} {et~al.}(1997){Chen}, {Shrader}, \& {Livio}}]{chen97}
{Chen}, W., {Shrader}, C.~R., \& {Livio}, M. 1997, \apj, 491, 312

\bibitem[{{Corral-Santana} {et~al.}(2011){Corral-Santana}, {Casares},
  {Shahbaz}, {Zurita}, {Mart{\'{\i}}nez-Pais}, \&
  {Rodr{\'{\i}}guez-Gil}}]{corralsantana11}
{Corral-Santana}, J.~M., {Casares}, J., {Shahbaz}, T., {Zurita}, C.,
  {Mart{\'{\i}}nez-Pais}, I.~G., \& {Rodr{\'{\i}}guez-Gil}, P. 2011, \mnras,
  413, L15

\bibitem[{{Curry} {et~al.}(2003){Curry}, {Bailyn}, \& {Buxton}}]{curry03}
{Curry}, S.~M., {Bailyn}, C., \& {Buxton}, M. 2003, in Bulletin of the American
  Astronomical Society, Vol.~35, American Astronomical Society Meeting
  Abstracts, 1331

\bibitem[{{della Valle} {et~al.}(1998){della Valle}, {Masetti}, \&
  {Bianchini}}]{dellavalle97}
{della Valle}, M., {Masetti}, N., \& {Bianchini}, A. 1998, \aap, 329, 606

\bibitem[{{Farr} {et~al.}(2011){Farr}, {Sravan}, {Cantrell}, {Kreidberg},
  {Bailyn}, {Mandel}, \& {Kalogera}}]{farr10}
{Farr}, W.~M., {Sravan}, N., {Cantrell}, A., {Kreidberg}, L., {Bailyn}, C.~D.,
  {Mandel}, I., \& {Kalogera}, V. 2011, \apj, 741, 103

\bibitem[{{Filippenko} {et~al.}(1999){Filippenko}, {Leonard}, {Matheson}, {Li},
  {Moran}, \& {Riess}}]{filippenko99}
{Filippenko}, A.~V., {Leonard}, D.~C., {Matheson}, T., {Li}, W., {Moran},
  E.~C., \& {Riess}, A.~G. 1999, \pasp, 111, 969

\bibitem[{{Filippenko} {et~al.}(1995){Filippenko}, {Matheson}, \&
  {Ho}}]{filippenko95}
{Filippenko}, A.~V., {Matheson}, T., \& {Ho}, L.~C. 1995, \apj, 455, 614

\bibitem[{{Filippenko} {et~al.}(1997){Filippenko}, {Matheson}, {Leonard},
  {Barth}, \& {van Dyk}}]{filippenko97}
{Filippenko}, A.~V., {Matheson}, T., {Leonard}, D.~C., {Barth}, A.~J., \& {van
  Dyk}, S.~D. 1997, \pasp, 109, 461

\bibitem[{{Frank} {et~al.}(2002){Frank}, {King}, \& {Raine}}]{fkr02}
{Frank}, J., {King}, A., \& {Raine}, D.~J. 2002, {Accretion Power in
  Astrophysics: Third Edition} ({Frank, J., King, A., \& Raine, D.~J.})

\bibitem[{{Froning} \& {Robinson}(2001)}]{froning01}
{Froning}, C.~S., \& {Robinson}, E.~L. 2001, \aj, 121, 2212

\bibitem[{{Fryer} \& {Kalogera}(2001)}]{fryer01}
{Fryer}, C.~L., \& {Kalogera}, V. 2001, \apj, 554, 548

\bibitem[{{Gelino} {et~al.}(2006){Gelino}, {Balman}, {K{\i}z{\i}lo{\u g}lu},
  {Y{\i}lmaz}, {Kalemci}, \& {Tomsick}}]{gelino06}
{Gelino}, D.~M., {Balman}, {\c S}., {K{\i}z{\i}lo{\u g}lu}, {\"U}.,
  {Y{\i}lmaz}, A., {Kalemci}, E., \& {Tomsick}, J.~A. 2006, \apj, 642, 438

\bibitem[{{Gelino} {et~al.}(2010){Gelino}, {Gelino}, \& {Harrison}}]{gelino10}
{Gelino}, D.~M., {Gelino}, C.~R., \& {Harrison}, T.~E. 2010, \apj, 718, 1

\bibitem[{{Gelino} \& {Harrison}(2003)}]{gelino03}
{Gelino}, D.~M., \& {Harrison}, T.~E. 2003, \apj, 599, 1254

\bibitem[{{Gelino} {et~al.}(2001){Gelino}, {Harrison}, \&
  {McNamara}}]{gelino01}
{Gelino}, D.~M., {Harrison}, T.~E., \& {McNamara}, B.~J. 2001, \aj, 122, 971

\bibitem[{{Gonz{\'a}lez Hern{\'a}ndez} {et~al.}(2008){Gonz{\'a}lez
  Hern{\'a}ndez}, {Rebolo}, {Israelian}, {Filippenko}, {Chornock}, {Tominaga},
  {Umeda}, \& {Nomoto}}]{gonzalezhernandez08}
{Gonz{\'a}lez Hern{\'a}ndez}, J.~I., {Rebolo}, R., {Israelian}, G.,
  {Filippenko}, A.~V., {Chornock}, R., {Tominaga}, N., {Umeda}, H., \&
  {Nomoto}, K. 2008, \apj, 679, 732

\bibitem[{{Goranskij}(1990)}]{goranskij90}
{Goranskij}, V.~P. 1990, Information Bulletin on Variable Stars, 3464, 1

\bibitem[{{Gray}(1992)}]{gray92}
{Gray}, D.~F. 1992, {The observation and analysis of stellar photospheres.}
  ({Gray, D.~F.})

\bibitem[{{Greene} {et~al.}(2001){Greene}, {Bailyn}, \& {Orosz}}]{greene01}
{Greene}, J., {Bailyn}, C.~D., \& {Orosz}, J.~A. 2001, \apj, 554, 1290

\bibitem[{{Greiner} {et~al.}(2001){Greiner}, {Cuby}, \&
  {McCaughrean}}]{greiner01}
{Greiner}, J., {Cuby}, J.~G., \& {McCaughrean}, M.~J. 2001, \nat, 414, 522

\bibitem[{{Harlaftis} {et~al.}(1999){Harlaftis}, {Collier}, {Horne}, \&
  {Filippenko}}]{harlaftis99}
{Harlaftis}, E., {Collier}, S., {Horne}, K., \& {Filippenko}, A.~V. 1999, \aap,
  341, 491

\bibitem[{{Harlaftis} \& {Greiner}(2004)}]{harlaftis04}
{Harlaftis}, E.~T., \& {Greiner}, J. 2004, \aap, 414, L13

\bibitem[{{Harlaftis} {et~al.}(1996){Harlaftis}, {Horne}, \&
  {Filippenko}}]{harlaftis96}
{Harlaftis}, E.~T., {Horne}, K., \& {Filippenko}, A.~V. 1996, \pasp, 108, 762

\bibitem[{{Harlaftis} {et~al.}(1997){Harlaftis}, {Steeghs}, {Horne}, \&
  {Filippenko}}]{harlaftis97}
{Harlaftis}, E.~T., {Steeghs}, D., {Horne}, K., \& {Filippenko}, A.~V. 1997,
  \aj, 114, 1170

\bibitem[{{Haswell} {et~al.}(1993){Haswell}, {Robinson}, {Horne}, {Stiening},
  \& {Abbott}}]{haswell93}
{Haswell}, C.~A., {Robinson}, E.~L., {Horne}, K., {Stiening}, R.~F., \&
  {Abbott}, T.~M.~C. 1993, \apj, 411, 802

\bibitem[{{Ioannou} {et~al.}(2004){Ioannou}, {Robinson}, {Welsh}, \&
  {Haswell}}]{ioannou04}
{Ioannou}, Z., {Robinson}, E.~L., {Welsh}, W.~F., \& {Haswell}, C.~A. 2004,
  \aj, 127, 481

\bibitem[{{Khargharia} {et~al.}(2010){Khargharia}, {Froning}, \&
  {Robinson}}]{khargharia10}
{Khargharia}, J., {Froning}, C.~S., \& {Robinson}, E.~L. 2010, \apj, 716, 1105

\bibitem[{{Lucy}(1967)}]{lucy67}
{Lucy}, L.~B. 1967, \zap, 65, 89

\bibitem[{{MacDonald} {et~al.}(2011){MacDonald}, {Bailyn}, \&
  {Cantrell}}]{macdonald11}
{MacDonald}, R.~K.~D., {Bailyn}, C.~D., \& {Cantrell}, A.~G. 2011, in Bulletin
  of the American Astronomical Society, Vol.~43, American Astronomical Society
  Meeting Abstracts \#217, \#144.20--+

\bibitem[{{Martin} {et~al.}(1995){Martin}, {Casares}, {Charles}, {van der
  Hooft}, \& {van Paradijs}}]{martin95}
{Martin}, A.~C., {Casares}, J., {Charles}, P.~A., {van der Hooft}, F., \& {van
  Paradijs}, J. 1995, \mnras, 274, L46

\bibitem[{{McClintock} {et~al.}(2001){McClintock}, {Haswell}, {Garcia},
  {Drake}, {Hynes}, {Marshall}, {Muno}, {Chaty}, {Garnavich}, {Groot}, {Lewin},
  {Mauche}, {Miller}, {Pooley}, {Shrader}, \& {Vrtilek}}]{mcclintock01}
{McClintock}, J.~E., {et~al.} 2001, \apj, 555, 477

\bibitem[{{Metropolis} {et~al.}(1953){Metropolis}, {Rosenbluth}, {Rosenbluth},
  {Teller}, \& {Teller}}]{metropolis53}
{Metropolis}, N., {Rosenbluth}, A.~W., {Rosenbluth}, M.~N., {Teller}, A.~H., \&
  {Teller}, E. 1953, \jcp, 21, 1087

\bibitem[{{Mirabel} \& {Rodr{\'{\i}}guez}(1994)}]{mirabelrodriguez94}
{Mirabel}, I.~F., \& {Rodr{\'{\i}}guez}, L.~F. 1994, \nat, 371, 46

\bibitem[{{Neilsen} {et~al.}(2008){Neilsen}, {Steeghs}, \&
  {Vrtilek}}]{neilsen08}
{Neilsen}, J., {Steeghs}, D., \& {Vrtilek}, S.~D. 2008, \mnras, 384, 849

\bibitem[{{Oke}(1977)}]{oke77}
{Oke}, J.~B. 1977, \apj, 217, 181

\bibitem[{{Orosz}(2003)}]{orosz03}
{Orosz}, J.~A. 2003, in IAU Symposium, Vol. 212, A Massive Star Odyssey: From
  Main Sequence to Supernova, ed. {K.~van der Hucht, A.~Herrero, \&
  C.~Esteban}, 365--+

\bibitem[{{Orosz} \& {Bailyn}(1995)}]{orosz95}
{Orosz}, J.~A., \& {Bailyn}, C.~D. 1995, \apjl, 446, L59+

\bibitem[{{Orosz} \& {Bailyn}(1997)}]{orosz97}
---. 1997, \apj, 477, 876

\bibitem[{{Orosz} {et~al.}(1996){Orosz}, {Bailyn}, {McClintock}, \&
  {Remillard}}]{orosz96}
{Orosz}, J.~A., {Bailyn}, C.~D., {McClintock}, J.~E., \& {Remillard}, R.~A.
  1996, \apj, 468, 380

\bibitem[{{Orosz} \& {Hauschildt}(2000)}]{orosz00}
{Orosz}, J.~A., \& {Hauschildt}, P.~H. 2000, \aap, 364, 265

\bibitem[{{Orosz} {et~al.}(1998){Orosz}, {Jain}, {Bailyn}, {McClintock}, \&
  {Remillard}}]{orosz98}
{Orosz}, J.~A., {Jain}, R.~K., {Bailyn}, C.~D., {McClintock}, J.~E., \&
  {Remillard}, R.~A. 1998, \apj, 499, 375

\bibitem[{{Orosz} {et~al.}(2004){Orosz}, {McClintock}, {Remillard}, \&
  {Corbel}}]{orosz04}
{Orosz}, J.~A., {McClintock}, J.~E., {Remillard}, R.~A., \& {Corbel}, S. 2004,
  \apj, 616, 376

\bibitem[{{Orosz} {et~al.}(2002){Orosz}, {Polisensky}, {Bailyn},
  {Tourtellotte}, {McClintock}, \& {Remillard}}]{orosz02}
{Orosz}, J.~A., {Polisensky}, E.~J., {Bailyn}, C.~D., {Tourtellotte}, S.~W.,
  {McClintock}, J.~E., \& {Remillard}, R.~A. 2002, in Bulletin of the American
  Astronomical Society, Vol.~34, American Astronomical Society Meeting
  Abstracts, 1124--+

\bibitem[{{Orosz} {et~al.}(2011){Orosz}, {Steiner}, {McClintock}, {Torres},
  {Remillard}, {Bailyn}, \& {Miller}}]{orosz11}
{Orosz}, J.~A., {Steiner}, J.~F., {McClintock}, J.~E., {Torres}, M.~A.~P.,
  {Remillard}, R.~A., {Bailyn}, C.~D., \& {Miller}, J.~M. 2011, \apj, 730, 75

\bibitem[{{Orosz} {et~al.}(2001){Orosz}, {Kuulkers}, {van der Klis},
  {McClintock}, {Garcia}, {Callanan}, {Bailyn}, {Jain}, \&
  {Remillard}}]{orosz01}
{Orosz}, J.~A., {et~al.} 2001, \apj, 555, 489

\bibitem[{{{\"O}zel} {et~al.}(2010){{\"O}zel}, {Psaltis}, {Narayan}, \&
  {McClintock}}]{ozel10}
{{\"O}zel}, F., {Psaltis}, D., {Narayan}, R., \& {McClintock}, J.~E. 2010,
  \apj, 725, 1918

\bibitem[{{Remillard} {et~al.}(1996{\natexlab{a}}){Remillard}, {Orosz},
  {McClintock}, \& {Bailyn}}]{remillard96}
{Remillard}, R.~A., {Orosz}, J.~A., {McClintock}, J.~E., \& {Bailyn}, C.~D.
  1996{\natexlab{a}}, \apj, 459, 226

\bibitem[{{Remillard} {et~al.}(1996{\natexlab{b}}){Remillard}, {Orosz},
  {McClintock}, \& {Bailyn}}]{remillard95}
---. 1996{\natexlab{b}}, \apj, 459, 226

\bibitem[{{Reynolds} {et~al.}(2007){Reynolds}, {Callanan}, \&
  {Filippenko}}]{reynolds07}
{Reynolds}, M.~T., {Callanan}, P.~J., \& {Filippenko}, A.~V. 2007, \mnras, 374,
  657

\bibitem[{{Reynolds} {et~al.}(2008){Reynolds}, {Callanan}, {Robinson}, \&
  {Froning}}]{reynolds08}
{Reynolds}, M.~T., {Callanan}, P.~J., {Robinson}, E.~L., \& {Froning}, C.~S.
  2008, \mnras, 387, 788

\bibitem[{{Sanwal} {et~al.}(1996){Sanwal}, {Robinson}, {Zhang}, {Colome},
  {Harvey}, {Ramseyer}, {Hellier}, \& {Wood}}]{sanwal96}
{Sanwal}, D., {Robinson}, E.~L., {Zhang}, E., {Colome}, C., {Harvey}, P.~M.,
  {Ramseyer}, T.~F., {Hellier}, C., \& {Wood}, J.~H. 1996, \apj, 460, 437

\bibitem[{{Shahbaz} {et~al.}(1994){Shahbaz}, {Naylor}, \&
  {Charles}}]{shahbaz94}
{Shahbaz}, T., {Naylor}, T., \& {Charles}, P.~A. 1994, \mnras, 268, 756

\bibitem[{{Shahbaz} {et~al.}(1999){Shahbaz}, {van der Hooft}, {Casares},
  {Charles}, \& {van Paradijs}}]{shahbaz99}
{Shahbaz}, T., {van der Hooft}, F., {Casares}, J., {Charles}, P.~A., \& {van
  Paradijs}, J. 1999, \mnras, 306, 89

\bibitem[{{Shahbaz} {et~al.}(1996){Shahbaz}, {van der Hooft}, {Charles},
  {Casares}, \& {van Paradijs}}]{shahbaz96}
{Shahbaz}, T., {van der Hooft}, F., {Charles}, P.~A., {Casares}, J., \& {van
  Paradijs}, J. 1996, \mnras, 282, L47

\bibitem[{{Timmer} \& {Koenig}(1995)}]{timmer95}
{Timmer}, J., \& {Koenig}, M. 1995, \aap, 300, 707

\bibitem[{{Valsecchi} {et~al.}(2010){Valsecchi}, {Glebbeek}, {Farr}, {Fragos},
  {Willems}, {Orosz}, {Liu}, \& {Kalogera}}]{valsecchi10}
{Valsecchi}, F., {Glebbeek}, E., {Farr}, W.~M., {Fragos}, T., {Willems}, B.,
  {Orosz}, J.~A., {Liu}, J., \& {Kalogera}, V. 2010, \nat, 468, 77

\bibitem[{{van der Hooft} {et~al.}(1998){van der Hooft}, {Heemskerk},
  {Alberts}, \& {van Paradijs}}]{vanderhooft98}
{van der Hooft}, F., {Heemskerk}, M.~H.~M., {Alberts}, F., \& {van Paradijs},
  J. 1998, \aap, 329, 538

\bibitem[{{van der Hooft} {et~al.}(1997){van der Hooft}, {Groot}, {Shahbaz},
  {Augusteijn}, {Casares}, {Dieters}, {Greenhill}, {Hill}, {Scheers}, {Naber},
  {de Jong}, {Charles}, \& {van Paradijs}}]{vanderhooft97}
{van der Hooft}, F., {et~al.} 1997, \mnras, 286, L43

\bibitem[{{Wagner} {et~al.}(2001){Wagner}, {Foltz}, {Shahbaz}, {Casares},
  {Charles}, {Starrfield}, \& {Hewett}}]{wagner01}
{Wagner}, R.~M., {Foltz}, C.~B., {Shahbaz}, T., {Casares}, J., {Charles},
  P.~A., {Starrfield}, S.~G., \& {Hewett}, P. 2001, \apj, 556, 42

\bibitem[{{Wagner} {et~al.}(1992){Wagner}, {Kreidl}, {Howell}, \&
  {Starrfield}}]{wagner92}
{Wagner}, R.~M., {Kreidl}, T.~J., {Howell}, S.~B., \& {Starrfield}, S.~G. 1992,
  \apjl, 401, L97

\bibitem[{{Webb} {et~al.}(2000){Webb}, {Naylor}, {Ioannou}, {Charles}, \&
  {Shahbaz}}]{webb00}
{Webb}, N.~A., {Naylor}, T., {Ioannou}, Z., {Charles}, P.~A., \& {Shahbaz}, T.
  2000, \mnras, 317, 528

\bibitem[{{Wong} {et~al.}(2012){Wong}, {Valsecchi}, {Fragos}, \&
  {Kalogera}}]{wong12}
{Wong}, T.-W., {Valsecchi}, F., {Fragos}, T., \& {Kalogera}, V. 2012, \apj,
  747, 111

\bibitem[{{Zurita} {et~al.}(2002){Zurita}, {S{\'a}nchez-Fern{\'a}ndez},
  {Casares}, {Charles}, {Abbott}, {Hakala}, {Rodr{\'{\i}}guez-Gil}, {Bernabei},
  {Piccioni}, {Guarnieri}, {Bartolini}, {Masetti}, {Shahbaz}, {Castro-Tirado},
  \& {Henden}}]{zurita02}
{Zurita}, C., {et~al.} 2002, \mnras, 334, 999

\end{thebibliography}
\bibliographystyle{apj}
\end{document}